\documentclass[12pt,preprint]{aastex}

\shorttitle{Observations and Simulations of Cygnus A}
\shortauthors{Carvalho et al.}

\begin{document}


\title{Comparison of Radio Observations and
Numerical Simulations of the Radio Lobes of Cygnus A}


\author{Joel C. Carvalho}
\affil{Space Telescope Science Institute, 3700 San Martin Dr.,
Baltimore, MD 21218}
\email{carvalho@dfte.ufrn.br}

\altaffiltext{1}{Departamento de Fisica, UFRN, C.P. 1661, CEP 59072-970, Natal,
RN, Brazil}

\author{Ruth A. Daly}
\affil{Department of Physics, Berks-Lehigh Valley College,\\ Penn State
University, Reading, PA 19610}
\email{rdaly@psu.edu}

\author{Matthew P. Mory}
\affil{Department of Physics, Berks-Lehigh Valley College,\\ Penn State
University, Reading, PA 19610}
\email{mmory@srtedg.com}

\and

\author{Christopher P. O'Dea}
\affil{Department of Physics, Rochester Institute of Technology, \\
 54 Lomb Memorial Drive, Rochester, NY 14623}
\email{odea@cis.rit.edu}



\clearpage

\begin{abstract}
We present a comparison of radio observations of the archetypal
powerful radio galaxy Cygnus A and 2-D numerical hydrodynamical simulations.
We characterize some global trends in the observed radio properties and
compare them with the properties of a simulated radio source.
The observational results are the following.
The width of the observed surface brightness distribution perpendicular to
the source axis can be well characterized by a Gaussian over most
of the length of the source. The ratio of the Gaussian FWHM to the
second moment is fairly constant along the source with an average
value of about 2.5 indicating that they give roughly consistent
 measurements of the source width. The average observed surface brightness,
estimated pressure, and estimated
minimum energy B field decrease with distance from the hot spots.
We find evidence for significant structure in the estimated
cross-sectional slices of emissivity. The numerical results are the
following.
Jets propagating in a constant density atmosphere will decelerate
with time. Thus, the estimated dynamical age of the source will be
greater than the actual age of
the source. For a source similar to Cygnus A the difference will
be about a factor of 2.
The second moment gives an accurate representation of the ``true"
width of the simulated source. The Gaussian FWHM tends to be about
40\% larger than the true width and can be systematically in error
if the surface brightness exhibits multiple peaks.
We suggest that the ratio of the Gaussian FWHM to the second
moment may be a diagnostic of the emissivity profile in the lobes.
The simulations can qualitatively reproduce the overall observed
morphology and the behavior of the cross-sections in surface
brightness, the decline in surface brightness with distance from
the hot spots, and the width of the lobes. This suggests that the
2-D simulations give a reasonable representation of the properties
of Cygnus A.
\end{abstract}


\keywords{galaxies: jets -- galaxies: intergalactic medium --
hydrodynamics -- radio continuum: galaxies}


\section{Introduction}

The primary physical processes in large, powerful FRII radio galaxies
are still poorly understood.  There is a consensus that the
sources are powered by two oppositely directed, highly collimated
outflows from an AGN - resulting in shocks and radio emission (Scheuer 1974;
Blandford \& Rees 1974).
It is clear that the youngest populations of relativistic electrons
are located near the radio hotspots at the extremities of
the source (e.g., Leahy, Muxlow, \& Stephens 1989;
Carilli et al. 1991; Alexander \& Leahy 1987).
Beyond this, differences of opinions begin to emerge.

It is important to understand the physical processes that shape
the observed radio emission of a source.  If these could
be identified and understood, it might be possible to use
the observed source properties to determine the pressure,
density, and temperature of the ambient gas, and the beam
power of the source.  These could then be used to probe
the evolution of source environments with redshift, and
evolution of AGN properties, such as jet power.  These, in turn,
would suggest the circumstances that lead to AGN activity,
and the role of environment in both triggering (perhaps via
mergers or close encounters) and then constraining (through the
gaseous environment) the radio properties of a source.

Much effort has been spent on analysis of multi-frequency radio
observations of radio galaxies (e.g., as summarized in Leahy 1991), and
independently on numerical simulations of radio galaxies
(e.g., Norman et al. 1982; Lind et al.
1989; Clarke, Norman \& Burns 1989; Hardee \& Norman 1990; Falle 1991;
Clarke \& Burns 1991; Cioffi \& Blondin 1992; Loken et al. 1992;
Mioduszewski, Hughes \& Duncan 1997; Marti et al. 1997;
Komissarov \& Falle 1997, 1998; Rosen et al. 1999; Carvalho \& O'Dea
2002a,b).

Here we attempt to bridge these two fields by directly comparing
radio images of the archetypal radio galaxy Cygnus A with the
results of 2-D hydrodynamical simulations of a source which is
expected to be similar to Cygnus A. Our goal is to explore the behavior
of global observed properties and relate these to intrinsic
properties of the underlying flows.

\section{Radio Observations \& Data Analysis}

Two observations of Cygnus A are studied in detail.  The
151 MHz radio image from
Leahy, Muxlow, \& Stephens (1989) was kindly provided to us
by Paddy Leahy, and the 1345 MHz image from Carilli et al. (1991),
was kindly provided to us by Chris Carilli.  The 151 MHz image,  obtained
with MERLIN, has an angular resolution of 3 arcsec down and mean and
rms noise levels of 0.1 and 0.3 Jy/beam.  The
1345 MHz image,  obtained with
the VLA, has an angular resolution of 1.36 arcsec down to
mean and rms noise levels of 0.01 and 0.01 Jy/beam.

\subsection{Surface Brightness Profiles and Gaussian Widths}

The FITS images of the source at 151 and 1345 MHz were rotated in an identical
manner so the hot spots lie on a horizontal line.
The hot spot-hot spot axis was taken to be a symmetry axis of the source.
Using AIPS and auxiliary programs,
vertical slices of surface brightness perpendicular to the hot spot symmetry axis
were taken at one pixel intervals of the FITS image, as illustrated
in Figure \ref{fig0}.
Each surface brightness profile was then studied in detail.  Distances
to each slice
are measured in arc seconds relative to the hot spot on each side
of the source, increasing as we move in toward the source center.

Each slice profile was fit with a Gaussian curve to
determine the FWHM of the best fitting Gaussian.
In these fits, we included the radio data above a threshold  of
  the mean plus three times the rms noise level of each image;
data below this level were removed.  These cuts were taken at
1 Jy/beam for the 151 MHz image  and 0.04 Jy/beam for the 1345 MHz
image. 

The radio surface brightness profiles and best fit Gaussian curves
are shown at integer intervals of
3 arc seconds (the FWHM of the 151 MHz image),
in Figures \ref{fig1L}, \ref{fig2L}, \ref{fig1R}, \ref{fig2R}
for the
151 and 1345 MHz data, and for each side of the source.

Generally, the Gaussian fits provide a good fit to the profiles;
occasionally the edges of the profiles drop more steeply
than the wings of the Gaussian fit.

The full width at half maximum of these Gaussian fits
are shown as a function of slice distance from the hot spot
at 151 and 1345 MHz in Figure \ref{fig3}.  The errors on the
Gaussian widths are much smaller than the FWHM of the radio
observations, so the one sigma error bar on the FWHM of the
Gaussian was taken to be half of the
FWHM of the observing beam.
Thus, an uncertainty of 1.5 arc seconds
for the 151 MHz data, and 0.7 arc seconds for the 1345 MHz data
was adopted as the uncertainty of the Gaussian widths.

The ratio of the best fit Gaussian widths at 151 and 1345 MHz
are shown in Figure \ref{fig4}.  The fact that this ratio
is unity over most of the source is reassuring, and suggests that
to first order the 151 and 1345 MHz data define similar
source shape and structure.  The average value of the ratio for the left and
right hand sides of the source, respectively, are $W_{151}/W_{1345} =
0.94 \pm 0.01$ and $0.90 \pm 0.01$ when all the data are included; when
only data with hot spot distances between 6'' and 48'' are included
these average value are $0.92 \pm 0.01$ and $1.01 \pm 0.01$ for
the left and right sides of the source, respectively.

\subsection{The First and Second Moments}

Two other measures of source structure are the first and second moments
of the cross-sectional surface brightness profiles described above.
These were obtained using the standard expressions
$\bar{x} = \sum(x_i~S_i)/\sum(S_i)$, and
$\sqrt{\sum(S_i~[x_i - \bar{x}]^2)/\sum(S_i)}$, where $S_i$ is the
surface brightness at the point $x_i$, shown as the
surface brightness and x-axis
in Figures \ref{fig1L} to \ref{fig2R}.
These are plotted as a function of the slice distance from the hot spot
in Figures \ref{fig5} and \ref{fig6}.  The uncertainty of the first moment is
taken to be half the beam width, as above.

The  first moment is a measure of the location of the surface brightness weighted
central axis of the source.  In the 151 and 1345 MHz data, we see
a varying displacement (or wandering) of the surface brightness weighted
source axis which is small
compared with the Gaussian FWHM of the bridge at similar bridge
locations.  The first moment
can be affected by the jet that is apparent in the radio grey scale
images, and it is interesting that the deviations of the first moment
from zero (expected for a perfectly cylindrically symmetric source) are in
opposite sense for the right and left-hand side of the source, which
could be influenced in part by any wobbling of the outflow axis of
the jets.  A detailed analysis indicates, however, that there is not an exact
correspondence between the deviations of the first moment on each side of the
source.

The second moment is another measure of the bridge width, and is
shown in Figure 9.  The bridge width as measured by the best fit
Gaussian and the second moment are compared in Figure \ref{width-ratio}, which
shows the ratio on the best fit Gaussian FWHM to the second moment.
For the 151 MHz data, this ratio is $2.59 \pm 0.07$ for the
left side of the source and $2.49 \pm 0.07$ for the
right side of the source, while
for the 1345 MHz data, this ratio is $2.45 \pm 0.03$ and
$2.53 \pm 0.03$ for the left and right hand sides
respectively.
The fact that the ratio is similar over most of
the source suggests that either method may be used to estimate the
bridge width. The physical interpretation of the numerical value of
the ratio is discussed in \S \ref{BW} (numerical simulations).
Note that the ratio $W_G/W_2$ discussed in \S 3.4,
obtained from the numerical simulations,
is half of the value stated here since $W_2$ is twice the
bridge radius indicated by the second moment.

\subsection{The Average Surface Brightness Along the Bridge}

The average surface brightness of each cross-sectional surface
brightness profile is shown in Figure \ref{fig8}.  This is obtained
by summing up
the total area under a particular surface brightness profile
(see Figures \ref{fig1L} to \ref{fig2R})
and dividing by the FWHM of the best fit Gaussian
curve, $W_G$: $\bar{S} = \sum(S_i \Delta x)/W_G$, where $S_i$ is
the surface brightness for the ith point, $\Delta x$ is the
interval along x around the point i.
Thus, the units of the average surface brightness are
Jy/beam.  Numerical simulations (see \S 3) indicate that
$W_G$ is a good estimate of the bridge diameter, suggesting that
the surface brightness values obtained and shown in
Figure \ref{fig8} do not need to be re-normalized.

The uncertainty of the average surface brightness is
obtained by adding the rms noise level of the image in quadrature
with the beam width: $(\delta \bar{S}/ \bar{S})^2 = (rms/ \bar{S})^2
+ (\theta_b/2W_G)^2$, where $\theta_b$ is the FWHM of the observing
beam; recall that the uncertainty on
the Gaussian width is taken to be half the beam size.

The contrast between the hot spots and the bridge is much higher
for the 1345 MHz data than it is for the 151 MHz data, and the values
of $\bar{S}$ differ substantially for the data sets.  But, once we
move beyond the hot spots into the bridge region, the
profile of the average surface brightness along the bridge axis
is similar for the two frequencies, which can be seen in Figure \ref{fig8}.

\subsection{The Average Pressure and Minimum Energy Magnetic Field
Along the Bridge}

The average bridge pressure for each slice is plotted as a function
of distance from the hot spot on each side of the source in Figure
\ref{fig9}.
This is a ``minimum energy'' pressure (Burbidge 1956)
in the sense that it is obtained using the equation
$P = (\bar{S}/W_G)^{4/7}$
(see Figures \ref{fig1L} to \ref{fig2R}).
Thus, the pressure has
units of (Jy/arc sec$)^{4/7}$.
To convert this to physical
units of erg/cm${}^3$ and thus obtain the normalization
factor to apply to the pressure presented, we note that
the pressure is simply related to the minimum energy
magnetic field (e.g. Wan et al. 2000):
$P = (1.33 b^{-1.5}+b^2)B^2/(24 \pi)$,
where $B$ is the minimum energy magnetic field strength and $b$
parameterizes the offset from minimum energy conditions: the
true field strength is $bB$.  As described below, the minimum
energy magnetic field strength estimated using the 151 MHz data
has a normalization factor of $30 \mu$G, so the normalization factor
for the pressure estimated using the 151 MHz data is about
$1.2 \times 10^{-11} \hbox{ erg~cm}^{-3}(1.33b^{-1.5}+b^2)$, or
about $2.8 \times 10^{-11} \hbox{~ erg cm}^{-3}$ for $b=1$, or
about $1.3 \times 10^{-10} \hbox{ erg cm}^{-3}$ for $b=0.25$, as
suggested by Carilli et al. (1991), and confirmed by
Wellman, Daly, \& Wan (1997a,b)
for the bridge region of Cygnus A.
At 1345 MHz the normalization factor for the pressure is
$10^{-10} \hbox{ erg cm}^{-3} (1.33 b^{-1.5}+b^2)$, or about
$2.4 \times 10^{-10} \hbox{ erg cm}^{-3}$ for $b=1$, or
about $1.1 \times 10^{-9} \hbox{ erg cm}^{-3}$ for $b=0.25$.

It is clear from Figure \ref{fig9} that the minimum energy magnetic
fields computed for the bridge region using the 151 MHz and 1345 MHz
data are similar in magnitude and structure,
while the pressures computed for the hotspots are
considerably different -  with the 1345 MHz data indicating a much higher
pressure than the 151 MHz data.

The average minimum energy magnetic field strength is each slice
is obtained by taking the square root of the minimum energy
pressure, $B =(\bar{S}/W_G)^{2/7}$.  The
normalization factor is obtained using the standard expression
for the minimum energy magnetic field strength, assuming zero
energy in relativistic protons, a filling factor of unity,
a constant spectral index of unity, 
frequency cut offs of 10 MHz and 100 GHz, and a cosmology
with a value of Hubble's constant of 70 km/s/Mpc, a normalized
mean mass density today of $\Omega_m=0.3$, and a cosmological constant
of $\Omega_{\Lambda}=0.7$.  These parameters yielded a normalization
for the minimum energy magnetic field strength for the
151 MHz data of about $30 \mu$G, and a normalization factor of about
$88 \mu$G for the 1345 MHz data.  The minimum energy magnetic field
strengths are shown in Figure \ref{fig10}.  Like the minimum energy
pressure, the minimum energy magnetic field strengths across the
bridge computed using the 151 MHz and 1345 MHz data are quite similar
in magnitude and structure, while the hotspot magnetic field
indicated by the 1345 MHz data is much larger than that indicated
by the 151 MHz data.  
One of the reasons that the estimated pressures in the radio hot spots
at 151 MHz and 1345 MHz differ is that the magnetic field strengths and
hence pressures have been estimated assuming 
a constant spectral index of unity.  While this may be a reasonable
estimate in the bridge region of the source, it clearly is not a 
good approximation in the hot spot region.  

Note that the total pressure $P$ in relativistic electrons and fields
estimated assuming that this pressure is
proportional to the square of the minimum energy magnetic field
implies that $P \propto (\epsilon)^{4/7}$, where $\epsilon$ is
the volume emissivity of the radio emission (see \S 2.5).  
This follows from the
fact that the radio surface brigthness $S$ is obtained by 
integrating the emissivity along the source path length, so  
$\bar{S} \sim \epsilon W_G$.  Thus, 
$P = (\bar{S}/W_G)^{4/7}$, which follows directly from the 
minimum energy argument, 
implies that $P \propto \epsilon^{4/7}$.
This is a good approximation whenever the true field strength is
proportional to the minimum energy magnetic field with a proportionality
constant that is independent of position in the source.  

\subsection{The Radio Emissivity as a Function of Position}

The emissivity, or energy emitted per unit volume per unit time
at one particular frequency can
be obtained from the surface brightness profiles shown in Figures
\ref{fig1L} to \ref{fig2R}.
For a particular surface brightness profile, it is assumed
that the emission is cylindrically symmetric, and that the
diameter of the cylinder is equal to the total width of that
profile (only data above the noise level of 1*mean + 3 rms
are included).  Then, the emissivity
as a function of radius from the center of the slice
can be determined.  This is done in the following way.
The data point that is furthest from the center of the slice
with a surface brightness above the noise level 
is used to determine the edge of the source.  We move in 
an amount $\Delta x$ to the position that is centered on 
$R_1$, and compute the emissivity 
$\epsilon_1$ at the point $R_1$ by taking the surface brightness at this
point divided by the length of the chord through the circle that
that point, as illustrated
in Figure \ref{fig13}.
We then have the emissivity $\epsilon_1$ at all points that
are at this distance, $R_1$, from the center of the cylinder for that slice.
Stepping in, the next surface brightness data point
can be used to determine the
emissivity $\epsilon_2$ that is a distance
$R_2$ from the center of the circle; this is done by accounting for the fact
that the total surface brightness $S_2$ will have contributions 
from $\epsilon_1$ and $\epsilon_2$, 
illustrated in Figure \ref{fig13}.  This process is continued to
obtain the emissivity as each distance $R_n$ from the center of the
axis of symmetry of the source, and is valid to the extent that
the slice possesses cylindrical symmetry.    

The normalization for the volume emissivity $\epsilon_{\nu} \equiv
dE/(dV~dt~d\nu)$ can be obtained by noting that
the emission coefficient $j_{\nu} = \epsilon_{\nu}/(4\pi)$ for
a volume element that emits isotropically, and the specific
intensity $I_{\nu} = j_{\nu} \Delta y$, where $\Delta y$ is the
length of the ray through the source that is shown in Figure \ref{fig13}.
The observed specific intensity $I_{\nu_o}=I_{\nu}(1+z)^{-3}$,
and has units of Jy/beam.  Using these expressions it
is easy to show that the normalization factor for the emissivities
shown in Figures \ref{figSBEML1}, \ref{figSBEML2},
\ref{figSBEMR1}, and \ref{figSBEMR2} are
$1.8 \times 10^{-34} \hbox{ erg s}^{-1} \hbox{ cm}^{-3} \hbox{ Hz}^{-1}$
for the 151 MHz data, and
$9.0 \times 10^{-34} \hbox{ erg s}^{-1} \hbox{ cm}^{-3} \hbox{ Hz}^{-1}$.

These emissivities show significant structure, indicating that in any given
slice there can be regions with very high and/or very low emissivities (or
that the assumption of cylindrical symmetry was not a valid assumption for
that slice).  A relatively large change in emissivity is needed to
produce a relatively small surface brightness feature.  Some of the
changes in emissivity are suggestive of a channel or jet-like feature,
which are suggested by either
an emission feature, or a cavity in the emissivity near the center of the
slice.  The variations in emissivity could be due to variations from
cylindrical symmetry, hydromagnetic waves, turbulence, or other
effects.

\section{Cygnus A {\it versus} Numerical Simulations}
\label{BW}

Here we present the results of 2-D numerical hydro simulations and
compare them with the observational results for Cygnus A. The
advantage is that in the simulations we know the {\it intrinsic\/}
properties of the ``source". The disadvantage is that the
simulations may not include all the relevant physics which occurs
in real radio galaxies.

\subsection{The Numerical Simulations}

We have carried out axisymmetrical hydrodynamical numerical
simulations of light jets propagating in a constant density
atmosphere. The details of the simulations are discussed by
Carvalho \& O'Dea (2002a,b). We have chosen not to include
magnetic fields in the simulations. If the magnetic field is 
dominant, MHD simulations produce structures
(e.g., the ``nose cone'', Clarke et al 1986; Lind et al. 1989) 
which do not appear to correspond to the observed radio 
structures in Cygnus A. If the magnetic field is not dominant,
the overall structure and dynamics of the source in the MHD
simulation are similar to the case where the magnetic field 
is absent . In addition, in a particle pressure
dominated source the magnetic field distribution closely follows
the distribution of gas pressure and density (Lind et al 1989). 
Numerical simulations by Komissarov (1989) show that if the
magnetic field is initially weak, it will come into equilibrium
with the gas pressure near the cocoon boundary because the
gas pressure drives the sideways expansion of the cocoon, and the
gas is not able to cross the magnetic field lines. 

Numerical simulations studies (e.g.,
Norman et al. 1982; Carvalho \& O'Dea 2002a,b; Krause 2003)
indicate that light jets will inflate a cocoon whose width is
inversely proportional to the density ratio ($\eta$) between the
jet and the ambient gas. The over expanded bridge in Cyg A implies
a very low value for $\eta$ ($<10^{-3}$). According to Alexander
and Pooley (1996) an order-of-magnitude estimate of the jet
density gives $\eta\sim 4\times 10^{-5}$. We have run a number of
simulations where we used $\eta = 2\times 10^{-4}$ and a jet
internal Mach number $M=10$ (Run D01), $\eta = 1\times 10^{-4}$,
$M=11$ (Run D02) and  $\eta = 1\times 10^{-4}$, $M=10$ (Run D03).
One of the reasons for doing so is to ensure that small changes on
the jet parameters do not have a significant effect on the final
results. Since the overall linear size of Cyg A is $\sim 120$ kpc
we used a simulation grid with physical dimension 66 kpc in the
direction of the source axis and a jet radius 1 kpc. We expect
that, as the bow shock in front of the jet head reaches the end of
the grid, the expanded cocoon would give a fair representation of
the source lobe.
We have run the simulations with several different grid widths
(equivalent to 16.5, 33 and 49.5 kpc) in order to study the effect
on the cocoon of the bow shock leaving the grid.
In the largest grid (Run D01), the entire source is contained on the grid,
while in the smaller grids, the bow shock eventually leaves the grid
laterally. We find that the details of the cocoon properties differ
in these simulations, though the overall properties of the cocoons
are similar. This gives us confidence in the results of simulations
carried out in the narrower  grids which are computationally much
cheaper.

The final stage of the simulation of Run D01 is shown as a density
contour map in Fig. \ref{ASUD_IndvPnlD01}. After comparing this
with Runs D02 and D03 we concluded that no significant change is
observed in the source physical properties although their general
appearance may look different. Therefore, in what follows we show
the results for Run D01 only. In Fig. \ref{ACygA-ViewD01}a we show
the contours of the bow shock, cocoon and jet. Fig.
\ref{ACygA-ViewD01} also shows the average pressure inside the
cocoon (b) the lateral expansion speed of the cocoon (measured at the
contact discontinuity) (c) and the lateral
expansion speed of the bow shock (measured
at the outer shock front) (d) along the source
axis.

The average pressure was obtained in the following way.
For each slice, with width equal to the grid cell and having
cylindrical symmetry, the total thermal energy was calculated
by summing up the product of the energy density per cell times
the volume element of the cell, and then dividing by the
total volume of the slice to obtain the total internal energy
density, $0.5 \rho v_{rms}^2$ or $3/2 nkT$.  
This was then multiplied by $(\Gamma - 1)$ 
to obtain the 
average pressure in that cell. The simulations was carried out for
a monotonic, ideal, 
non-relativistic fluid, thus ratio of specific heats is $\Gamma = 5/3$.

The shock structure seen in the numerical simulation is shown in
Figure \ref{ACygA-ViewD01}a.  As usual, three shock regions can be identified: the
outer shock front, the contact discontinuity, and the inner shock,
which includes the entire bridge region.  

In the simulation, the location of the jet can also be identified.
The outer shock front is also called the bow shock, and is
indicated by the outermost line in Figure 20a. The contact
discontinuity separates the shocked ambient gas from the shocked
jet fluid and is indicated by the middle line in Figure 20a.  The
radio emitting fluid that would produce the observed radio
emission is expected to lie within the contact discontinuity.
Thus, the region interior to the middle line in Figure 20a is what
is referred to as the ``cocoon'' throughout this paper.  The
shocked ambient gas lies between the bow shock and the cocoon and
is not discussed further in this paper.
The jet is identified in the simulations by means of trace
particles which are injected together with the jet gas. 

Slices are taken across the cocoon, and the average pressure
within the cocoon is computed for each slice as described above.  
This average pressure is plotted as a function
of distance from the hot spot in Figure 20b, and is normalized to
be unity at the location of the hot spot.  This measure of the
pressure includes only the thermal pressure of the ``gas'' within
the cocoon of the numerical simulation and does not include
kinetic pressure.  In the numerical simulation, the average
pressure in slices across the cocoon undergo a series of
oscillations as we move along the symmetry axis of the bridge.

The lateral expansion of the cocoon (i.e.,  as defined by
the contact discontinuity)
and that of the bow shock are shown in Figures \ref{ACygA-ViewD01}c and d,
and are normalized to be unity at the location of the hot spot.
Like the bridge width, and the average pressure in a slice across
the bridge (i.e., the cocoon), the lateral expansion speed of the
cocoon undergoes a series of oscillations rather than varying smoothly
along the contact discontinuity.  This variation in lateral expansion
speed is reflected in differences in the bridge width.  For example,
in the simulation the
bridge width is seen to have a bulge at distances from the hot
spot of about 10 to 30 kpc, and this corresponds to a region
where there is an increase in the lateral expansion speed
of the cocoon relative to nearby points.  The lateral expansion
speed of the bow shock varies much more smoothly and monotonically with
distance from the hot spot.  And, appears to generally follow
$U_b \propto D^{1/2}$, which is predicted theoretically for
the regions in which the
expansion speed exceeds the ambient sound speed, after which the
expansion speed is expected to level off (e.g., Begelman \& Cioffi
1989; Daly 1990, 1994; Wellman et al. 1997a; Reynolds, Heinz,
\& Begelman 2002; Carvalho \& O'Dea 2002a).    

\subsection{Dynamical and Synchrotron Age}

The ratio between the source advance speed and the initial jet
speed is shown in Fig. \ref{AHeadSpeedD01}  as a function of
source size. For comparison, we also show the ambient sound speed
(dashed line). We see how the source decelerates rapidly with its
Mach number relative to the ambient sound speed decreasing by a
factor of $\sim 22$ from $\sim 70$ to $\sim 3.2$. This is due to
the progressive increase of the jet head radius. The observed
speed fluctuations are also due to oscillations of the head
radius. These oscillations of the head radius are probably driven
by vortex shedding (e.g., Lind et al 1989; Norman, Winkler, \&
Smarr 1982). We note that changes in jet momentum flux may also
drive changes in the advance speed. Near the end of the 
simulation, an increase is observed
in the advance speed which experience indicates is a transient
phenomenon.

Two effects likely to occur in nature that are not included in
the simulation may also come into play.  In nature, the density of
the ambient gas may be slowly decreasing with distance from the
central engine rather than remaining constant, and this would
cause the decline in the head velocity to be less significant.
On the other hand, in nature the outflow axis of the jet may ``wobble''
rather than pointing in a fixed direction (as in the simulation),
which would go in the opposite sense, and at any given time would
cause the velocity of the head to be lower (e.g. Scheuer 1982; 
Cox, Gull, \& Scheuer 1991).   
This second effect could
also modify the bridge width as a function of distance from
the hot spot. 

The dynamical age of Cygnus A determined using the ram pressure of
the hotspot is about an order of magnitude larger than the
estimated synchrotron age (Carilli et al. 1991). It has been
suggested by some authors that this discrepancy can be alleviated
if one relaxes the equipartition hypothesis when estimating the
magnetic field. According to Carilli et al. the magnetic field
must be 1/3 of the equipartition value. Similar results have been
found by Perley and Taylor (1991) for 3C295, and by
Wellman, Daly and Wan (1997a,b) for a large sample
of sources.  The deceleration of the source head observed in
the simulations can, at least in part, be responsible for the
discrepancy.

Thus, the age calculated using the speed at the source maximum length is
larger than the real age of the source. In our simulation the
difference is not too large because at the end of the simulation
the head speed has momentarily increased. However, if we calculate
the age by taking the head speed when the source was, for
instance, $\sim 51$ kpc in size, it would be larger than the real
age by a factor $\sim 2$.
Thus, this effect can account for only a part of the discrepancy
between the ram-pressure balance and synchrotron ages found by
Carilli et al. (1991).

\subsection{Surface Brightness and Radio Emissivity}

We have calculated the volume emissivity of the source assuming
that the energy density of the relativistic electron component is
proportional to the hydrodynamical (thermal)
pressure of the gas inside the
cocoon and an equipartition magnetic field.
This is easy to do since the radio emissivity is
proportional to $P^{7/4}$, as described at the end of \S 2.4. 
We will also consider
separately contributions proportional to
the thermal pressure, the kinetic pressure, and the total pressure
(see Figure 25).
We assume that the
radio emission comes from the jet itself and the cocoon (Leahy,
1991) where the pressure is higher than ambient pressure. This
region (corresponding to the region interior to the middle line
in Figure 20a) 
is filled with jet material which has been shocked in the
head and expands laterally. Besides any non-relativistic gas that
may be present in this region, it is assumed that the region also
contains the basic ingredients necessary to generate synchrotron
emission, that is, a mixture of magnetic field and relativistic
electrons most probably  accelerated by Fermi process at the
strong shock region in the jet head. We integrated the radio
emissivity along the line of sight (assuming the source lies
in the plane of the sky)
to obtain the surface brightness, which allows us to
draw synthetic radio images of the source. We have also taken into
account radiation loss of the relativistic electron population in
a very simple way. We divided the cocoon into slices and assigned an
age to each slice equal to the time that has passed since it has
been laid down by the advancing jet head. The modified spectrum is
then calculated using the Kardashev-Pacholczyk model (Kardashev 1962;
Pacholczyk 1970) supposing that the magnetic field is the 
equipartition field.

In this manner, we were able to predict the radio structure
that would be observed from the source indicated by the numerical
simulation.
In Fig. \ref{AradioD01te} we show the predicted 151 MHz radio image
of the source at five evolutionary stages; the synthetic data have 
been smoothed by a Gaussian with FWHM of 3 arcsec to match the 151 MHz data, 
as shown by the black circle in panel e of the figure.  
The age of the source corresponding to each panel is: (a) 1.35 Myr, 
(b) 4.53 Myr, (c)
6.17 Myr, (d) 8.89 Myr and (e) 9.15 Myr. One can clearly see the
characteristic edge-brightened shape of powerful FR-II sources and
an aspect ratio resembling that of Cygnus A. 
When the source is
young, radiative energy losses are negligible and we can see the
radio bridge extending all the way from the head to the central
galaxy (panels b and c). As the source ages, the central parts
becomes less luminous and it assumes the classical double lobe
shape (panel e). The hotspots are evident in all the simulated images.

Cross-sectional slices of the surface brightness distribution
for the source as it appears in panel e, when it is about 9 Myr old, 
are shown in Fig. \ref{ACygA-SBD01te}. The second panel in the first row
corresponds to the position of the hotspot while the bottom ones
show the parts of the source near the central object. We have
normalized the surface brightness to unity at the hot spot and the
scale in each row is the same as that in Fig. \ref{fig1R} at 151
MHz. The simulated surface brightness is qualitatively similar to
that observed in Cygnus A (Figs. \ref{fig1L}, \ref{fig2L},
\ref{fig1R}, \ref{fig2R}).  Of course, the simulation is axisymmetric
and thus so are the cross-sectional surface brightness slices, whereas
in nature this is not always the case.

The Gaussian fit is shown as a dotted line in the figure.
The central spikes that appear in some of the panels are caused by the jet
contribution to the surface brightness.  The spikes are more evident in
those regions of the jet where it has been compressed
(Fig. \ref{ASUD_IndvPnlD01}).  The Gaussian fits provide a reasonable
approximation to the actual surface brightness slices over most
of the source except along row 3 where the wings of the Gaussian fit clearly
extend well beyond the boundary of the source.

Fig. \ref{ACyg_S_toteki} shows the predicted average surface brightness of a
slice as a function of distance from the hotspot in the
numerical simulation; the average surface brightness is obtained
as described in section 2.3, but using the true width of the 
simulated source.  

Here we investigate how the different forms of energy (thermal
pressure, kinetic pressure, and total pressure) could contribute
to the radio emission. Again, the pressure in relativistic
electrons and magnetic fields is assumed to be proportional to the
gas pressure, this being determined by
the local thermal pressure, kinetic pressure, or total pressure
given by the numerical simulation. The thermal pressure is the
thermodynamic pressure associated with the gas microscopic motion
(discussed in \S 3.1).  
The kinetic pressure is equal to $(\Gamma -1)$ times the kinetic
energy density, where the kinetic energy density 
$0.5 \rho v^2$ depends on the large-scale, macroscopic or 
bulk velocity $v$ of the gas.  
The total pressure is the sum of the thermal and kinetic
pressures. We also use the approximation  that the emissivity is
proportional to $P^{7/4}$ (described in \S 2.4).  
Considering the kinetic pressure allows for the fact that macroscopic
kinetic energy in the form of turbulent motion could be
transferred to the relativistic component through the Fermi
process or through hydromagnetic instabilities (e.g. De Young 1980;
Eilek \& Henriksen 1984; Eilek \& Shore 1989).  
For this figure,
the radio emissivity is calculated supposing that the local  
pressure of the relativistic fluid, including the 
relativistic electrons and magnetic fields,  
is equal to the total pressure (full
line), thermal pressure (dashed-dotted line) and the kinetic
pressure (dotted line).

We observe that in many
places the contribution of the kinetic pressure can be very
important. This is a feature that is not being considered by the
majority of works in the literature.
It is very interesting to note that the  surface brightness
profile indicated by the total pressure (kinetic plus thermal)
is very similar to that of the 151MHz data (see Figure 11).
The surface brightness profile indicated by the thermal pressure alone
is very similar to that of the 1345 MHz data (see Figure 11).
This suggests that different forms of energy within the bridge
may enhance radio emission at different frequencies differently.
It is easy to see how this could occur since the radio emission
at different frequencies is produced by relativistic electrons
with different Lorentz factors.  Thus, it would only require that
the thermal pressure produces an underlying relativistic electron
population with a power law distribution at all Lorentz factors,
while the kinetic pressure boosts additional relativistic electrons
to Lorentz factors that produce the 151 MHz emission but 
not the 1345 MHz emission.  For example, through hydromagnetic
processes the kinetic pressure could boost relativistic electrons
with lower Lorentz factor up to the Lorentz factors needed to
produce 151 MHz emission.  This could help to explain some of
the changes in the radio spectrum seen across the source.

To investigate this further,  we compare the average thermal pressure, 
and the total average 
pressure (thermal plus kinetic pressures) 
obtained from the numerical simulation with the average pressure
obtained from the 151 MHz and 1345 MHz data (Figure 
\ref{ACyg_PressureAll}).  To do this, the physical value of
the units on the pressure
obtained from the simulation were investigated, and were determined 
to be $1.596\times 10^{-8} n_a \hbox{ dyn/cm}^2$, where
$n_a$ is the ambient gas density.   
A value of $n_a = 10^{-2}~\hbox{cm}^{-3}$ was adopted
since this is consistent with X-ray measurements of the 
gas in the vicinity of the radio source (e.g., Smith et al. 2002).  
The behavior of the pressures in the simulations roughly 
tracks that seen in the data.  Of course, it is possible that
the value of $b$ in the hot spot region is closer to unity 
(e.g., Hardcastle et al. 2004; Donahue, Daly, \& Horner 2003; 
Hardcastle et al. 2002; Wilson, Young, \& Shopbell 2000),
while that in the bridge is $\sim 0.25$.  (e.g., Carilli et al. 1991;
Perley \& Taylor 1991; Wellman et al. 1997b).  
The value of $b$ does not affect the 
pressures determined using the numerical code (which only
depend on the scaling factor $n_a$), but it does affect
the pressure determined from the data.  If b in the hot spot
in closer to unity, then the pressure determined using the data
will be lower in this region that the values shown.  This 
would improve the fit to the 1345 MHz data, and worsen the match
with the 151 MHz data. Another effect to consider is 
the efficiency factor (converting gas kinetic into relativistic
energy) used in converting surface brightness into pressure, which
could vary with location within the source or could
have an energy dependence, 
resulting in a frequency dependence. In addition, 
consider particle re-acceleration within the bridge region or a
backflow could be important in the source.  
In view of these factors, the rough overall agreement between the 
data and the simulation is encouraging.

Finally, in Fig. \ref{ACyg_Pressure_var} we compare the thermal pressure
determined from the simulation at several different times with 
that  estimated from the 151 MHz empirical data
for the two sides of Cygnus A, each normalized to match at the 
peak of the hot spot.  
Sources are
dynamic, constantly changing with time, so we consider snapshot
corresponding to panels c, d, and e of Figure 24 re-scaled to
have a length of 60 kpc.
The dynamic nature of the source
is obvious from these snapshots.
The agreement is better than for the
surface brightness distribution and this is due to the fact that
the simulated pressure profile does not depend upon the bridge
width. As we shall see below, the simulated and observed width do
not agree equally well over the entire bridge extent.
We see that the pressure profile varies considerably with time
over the course of the simulation. This implies that there is
no ``steady-state" pressure distribution in the lobes. 

\subsection{Bridge Width}

We compare the ``true" width of the bridge in the simulation 
(Figure~\ref{ACygA-ViewD01}a) with the FWHM and second 
moment of the synthetic surface brightness distribution 
(Figure \ref{ACygA-BridgeWD01}), where only the thermal component
of the surface brightness was considered. 
We find that while
the second moment gives a good fit to the true cocoon width, the
Gaussian FWHM deviates significantly. (Note: this is not seen
in the data, where the second moment and FWHM of the Gaussian
fits track each other quite well Figure~\ref{width-ratio}). 
The main departure in the simulation occurs
between 20 and 30 kpc from the hot spot.
This can be attributed to the fact that the
cross-sectional slices of the surface brightness distribution in
this region exhibits a double peak as can be seen in Fig.
\ref{ACygA-SBD01te}. This double peak implies that the gas pressure
distribution inside the simulated cocoon is higher near its edges,
at least in the region in question ($20 - 30$ kpc).

The ratio $\zeta=W_G/W_2$ between the two quantities is 
in the range $\sim 1.2 - 1.8$
with an average value $1.39\pm 0.07$. For a perfect Gaussian
distribution, the ratio $\zeta=\sqrt{2\ln 2}\simeq 1.177$. To
compare this with the empirical data shown in Fig. \ref{width-ratio} we
note that in this figure we have plotted  $2\zeta =
W_G/$(second moment). Since the value of the empirical data is in
the range $2 - 3$ we conclude that the results of the simulations
$2\zeta \sim 2.4 - 3.6$ are in fair agreement with the data.

Figure \ref{ACyg_Width} shows a comparison between the observed FWHM
of the radio source with the ``true" width of the simulated 
source. We see that the simulation shows the same general behavior
as the radio source.

\subsubsection{Emissivity Distribution}

As mentioned above, the value of ratio $\zeta=W_G/W_2$ will
depend on the emissivity distribution across the cocoon. In this
sense, the ratio may be used as a diagnostic of the
distribution of emissivity.

We have thus investigated this by  calculating the ratio
$\zeta$ for a simple model cocoon whose radius varies smoothly
according to
$$
R_c = 0.5 + 9.5 \left({D\over 60kpc}\right)^{1/2}
$$
where $D$ is the distance from the hot
spot and $R_c$ is in kpc. We studied five cases with different
pressure profiles
$P(r)$. The radial dependence of $P$ is shown
in Table \ref{zeta}. In all cases $P = 0 $ for $r>R_c$.

We calculated the surface brightness using the same beam size as
the empirical data and calculated the second
moment and fit a Gaussian to the surface brightness profile
perpendicular to the cylinder axis in order to determine the ratio
$\zeta$. The values of  $W$, $W_G$ and $W_2$ for four of the model bridges in
Table \ref{zeta} are shown in Fig. \ref{ACyg_Width4}. We can
clearly see how the relation between the true width $W$ and the
calculated ones $W_G$ and $W_2$ depends strongly on the pressure
distribution across the cocoon.

In Table \ref{zeta} we show the average values of the Gaussian
width, the second moment and $\zeta$ for the five distributions.
We see that for Model A both the Gaussian fit and the second
moment are about 55\% to 70\% of the true width, and thus need
to be normalized to be used as a direct
measure of the bridge width. For the other models the
Gaussian width increasingly departs from the true width as the
peak of pressure distribution moves away from the source axis. On
the other hand, the second moment gives a better estimate when
the pressure distribution concentrates near the border of the
cocoon.

We notice that the range of values of the ratio $\zeta$ of
the Gaussian FWHM and the second moment width $W_2$ for the five
distributions is in agreement with the value from the numerical
simulations ($1.2 - 1.8$). The average value observed in Cygnus A
($\sim 2.5=2\times 1.25$) is compatible with an emissivity distribution
concentrated near the source axis (Model A).
In a few cases when it
reaches higher values ($\sim 3=2\times 1.5$), the pressure distribution
in these regions must have two separate peaks as in Model E.
In fact, if we look at the deconvolved emissivity distribution of
Cygnus A at 151 MHz (Figs.
 \ref{figSBEML1}, \ref{figSBEML2}, \ref{figSBEMR1}, \ref{figSBEMR2})
we see that this seems indeed to be the case in several sections
of the bridge. Thus, the ratio between the
Gaussian width and the second moment can provide a diagnostic of
the emissivity profile in the lobe. It may be necessary to run 3-D
numerical simulations to properly ``calibrate" this diagnostic
ratio.

We have seen that, for the numerical simulation,
the second moment of the surface brightness
distribution gives a more stable estimate of the bridge width in
our model cocoon than the Gaussian fit (e.g., Fig. \ref{ACygA-BridgeWD01}).
If this applies to real sources it can be used as a measure of its
real width, provided we have a proper normalization factor. 
Our simulation gives an average value for the ratio
between the cocoon and Gaussian width $\beta_G = W/W_G$
of $\beta_G = 0.91 \pm 0.12$; while the average ratio of the
true width to the second moment $\beta_2 = W/W_2$ 
is $\beta_2 = 1.26 \pm 0.04$. 

We note that (using $\zeta=W_G/W_2$ as a diagnostic, Table \ref{zeta})
the range of values of $\zeta$ found in Cygnus A indicates that,
at least in some regions, the bridge contains high pressure edges.
Carilli et al. (1991) suggest that an edge brightening of the lobes
could account for the increase in the magnetic field from the center
to the edge of the
lobes needed to explain the estimated age gradient. If this is the
case, the Gaussian fit overestimates the bridge width as in Models
D and E (Table \ref{zeta}) and, again, the second moment should be
used to calculate the bridge width.

\section{SUMMARY}

We present a comparison of radio observations of the archetypal
powerful radio galaxy Cygnus A and 2-D numerical hydrodynamical simulations.
We characterize some global trends in the observed radio properties and
compare them with the properties of a simulated radio source.
We have analyzed detailed radio images at 151 MHz (Leahy
et al. 1989) and 1345 MHz (Carilli et al. 1991). We have determined the
surface brightness, emissivity, Gaussian FWHM, first and second moment of
cross-sectional slices of the source as a function of distance along the
bridges.  We have also estimated the average pressure and minimum
energy magnetic field in the radio plasma as a function of distance
along the bridges. We find the following trends in global properties.

\begin{itemize}
\item
The first moment of the brightness distribution perpendicular to
the source axis is a measure of the surface brightness weighted
center of the source. We find that the first moment wanders with
a peak-to-peak amplitude of $\sim 10$ arcsec. The sense of the
wander has the opposite sign on the two sides of the source,
but there is not an exact correspondence.

\item
The width of the surface brightness distribution perpendicular to
the source axis can be well characterized by a Gaussian over most
of the length of the source. The ratio of the Gaussian FWHM to the
second moment is fairly constant along the source with an average
value of about 2.5 indicating that they give roughly consistent
 measurements of the source width. The radio lobe increases in width with
distance from the hot spot. Superimposed on this general expansion
are several ``wiggles."

\item
The average surface brightness, estimated pressure, and estimated
minimum energy B field decrease with distance from the hot spots.

\item
We find evidence for significant structure in the estimated
cross-sectional slices of emissivity.
\end{itemize}

Following the methods of Carvalho \& O'Dea (2002a,b) we present the results
of 2-D numerical hydrodynamical simulations of light (density contrast
$\zeta = 10^{-4}$), supersonic (jet Mach number $M= 10$)  jets propagating
in a constant density environment.
The results from the simulations are the following.

\begin{itemize}
\item Jets propagating in a constant density atmosphere will decelerate
with time (see also Carvalho \& O'Dea 2002a). Thus, the estimated
dynamical age of the source will be greater than the actual age of
the source. For a source similar to Cygnus A the difference will
be about a factor of 2.

\item We have run the simulations with several different grid widths.
In the largest grid, the entire source is contained on the grid,
while in the smaller grids, the bow shock eventually leaves the grid
laterally. We find that the details of the cocoon properties differ
in these simulations, though the overall properties of the cocoons
are similar.

\item
The pressure profile  varies considerably with time
over the course of the simulation. This implies that there is
no ``steady-state" pressure distribution in the lobes. 
There is considerable time dependence in the variation of the
average surface brightness in the lobe near the hot spots,
possibly due to vortex shedding.

\item The kinetic pressure in the cocoon (due to large scale turbulence)
can make a significant contribution to the total pressure of the cocoon.
We speculate that
the thermal pressure produces an underlying relativistic electron
population with a power law distribution at all Lorentz factors,
while the kinetic pressure adds ``extra'' relativistic electrons
with the Lorentz factors that produce the 151 MHz emission and
not the 1345 MHz emission.  For example, through hydromagnetic
processes the kinetic pressure could boost relativistic electrons
with lower Lorentz factor up to the Lorentz factors needed to
produce 151 MHz emission.  This could help to explain some of
the changes in the radio spectrum seen across the source.

\item
The second moment gives an accurate representation of the ``true"
width of the simulated source. The Gaussian FWHM tends to be about
40\% larger than the true width and can be systematically in error
if the surface brightness exhibits multiple peaks.
We suggest that the ratio of the Gaussian FWHM to the second
moment may be a diagnostic of the emissivity profile in the lobes.
However, ``calibration" of this diagnostic probably requires
3-D simulations. Applying this diagnostic reveals that bridge contains 
high pressure edges. This is consistent with the 
suggestion of Carilli et al. (1991) that an edge brightening of the lobes
could account for the increase in the magnetic field from the center
to the edge of the lobes needed to explain the estimated age gradient.
\end{itemize}

We compare our simulations with the properties of the observed radio
sources in two complementary ways. We note that the pressure is
directly determined in the simulation but is estimated from the radio
data, while the surface brightness is directly determined from the
radio data but is estimated from the simulations.

The simulations can qualitatively reproduce the overall observed
morphology and the behavior of the cross-sections in surface
brightness, the decline in surface brightness with distance from
the hot spots, and the width of the lobes. This suggests that the
2-D simulations give a reasonable representation of the properties
of Cygnus A.

This is especially encouraging given that the outflow axis of the jets
in Cygnus A may wobble while those in the numerical simulation are held
steady, and given that the ambient gas density in the simulation is
assumed to be constant.  The good agreement between the simulation
and the data suggest that these effects are relatively small for
a source like Cygnus A.

\acknowledgments

We are grateful to Paddy Leahy and Chris Carilli for sending us their
radio data and to the referee for a thoughtful review of the manuscript.  
JCC acknowledges the financial support of PRONEX/Finep and CNPq and the
hospitality of STScI where this work was carried out.
This work was supported in part
by a grant to C. O'Dea from the STScI Collaborative
Visitor Program which funded visits by JCC to STScI, and by
grant  AST-0206002 to R. Daly from
the U. S. National Science Foundation.



\acknowledgements

\begin{deluxetable}{lcccc}
\tablecolumns{5}
\tablewidth{0pc}
\tablecaption{Comparison between calculated and true width for a model bridge\label{zeta}}
\tablehead{
&\colhead{Pressure}    &\colhead{Gaussian fit}    &\colhead{Second moment}&\\
&\colhead{distribution}&\colhead{$<W_G/2R_c>$}&\colhead{$<W_2/2R_c>$}&\colhead{$<\zeta=W_G/W_2>$}\\
&\colhead{$P(r)$}&&&\\}
\startdata
A&$ 1- {r\over R_c}$                & 0.673  &  0.550 & 1.223 \\
B&       1                          & 1.075  &  0.786 & 1.367 \\
C&$\left({r\over R_c}\right)^{1/2}$ & 1.211  &  0.848 & 1.428 \\
D&$\left({r\over R_c}\right)^1$     & 1.301  &  0.889 & 1.463 \\
E&$\left({r\over R_c}\right)^{2}$   & 1.416  &  0.939 & 1.507 \\
\enddata
\end{deluxetable}

\clearpage

\begin{figure}
\caption{Examples of cross-sectional slices used to define
the surface brightness profiles shown in Fig. 2 for the
151 MHz data (left) and the 1345 MHz data (right).
The distance of the slice from the hot spot is indicated
at the bottom on the slice.  The images have been rotated slightly for
ease of display and the left side is the south-east lobe and the right side
is the north-west lobe.
\label{fig0}}
\end{figure}

\clearpage

\begin{figure}
\plottwo{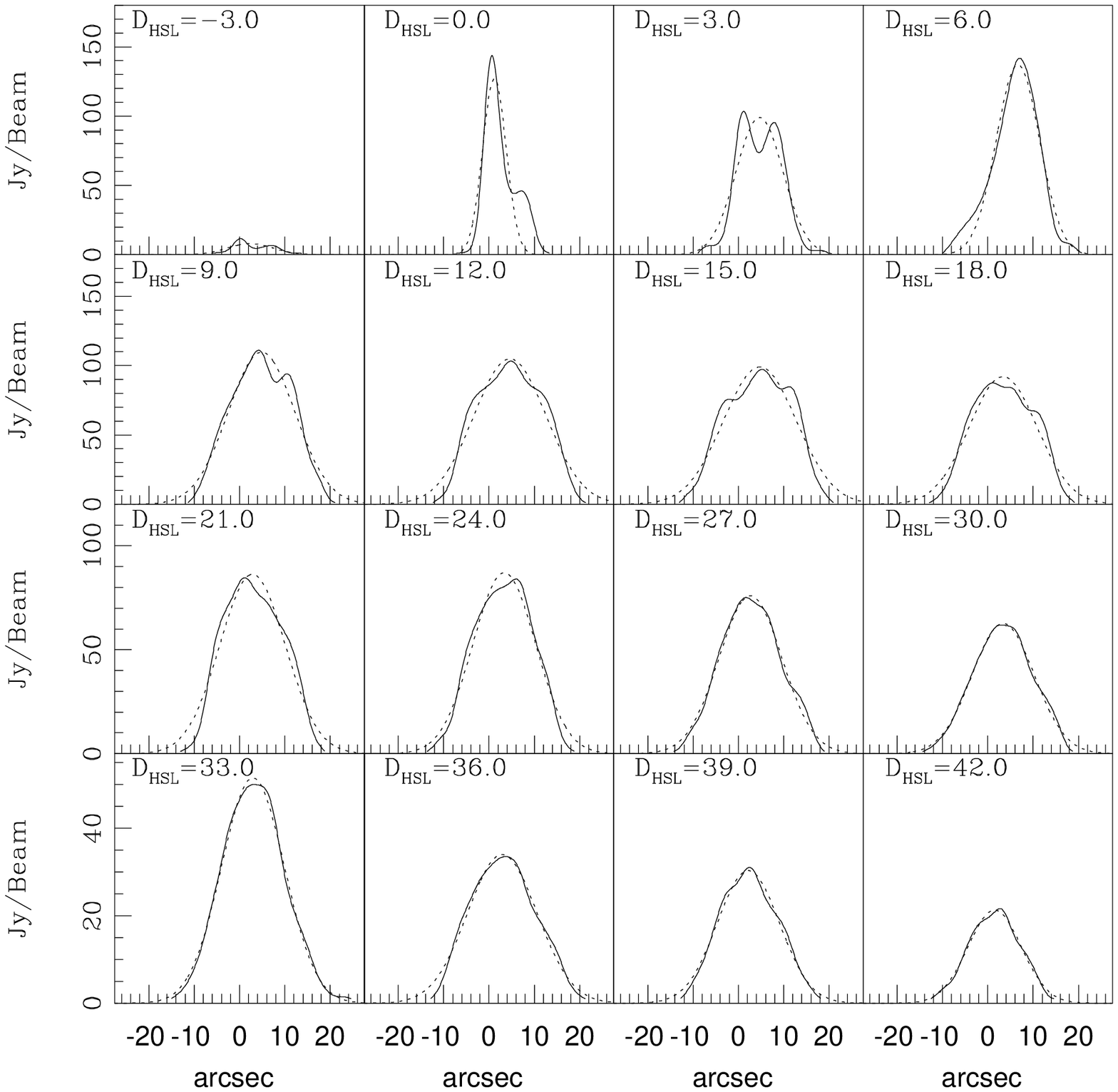}{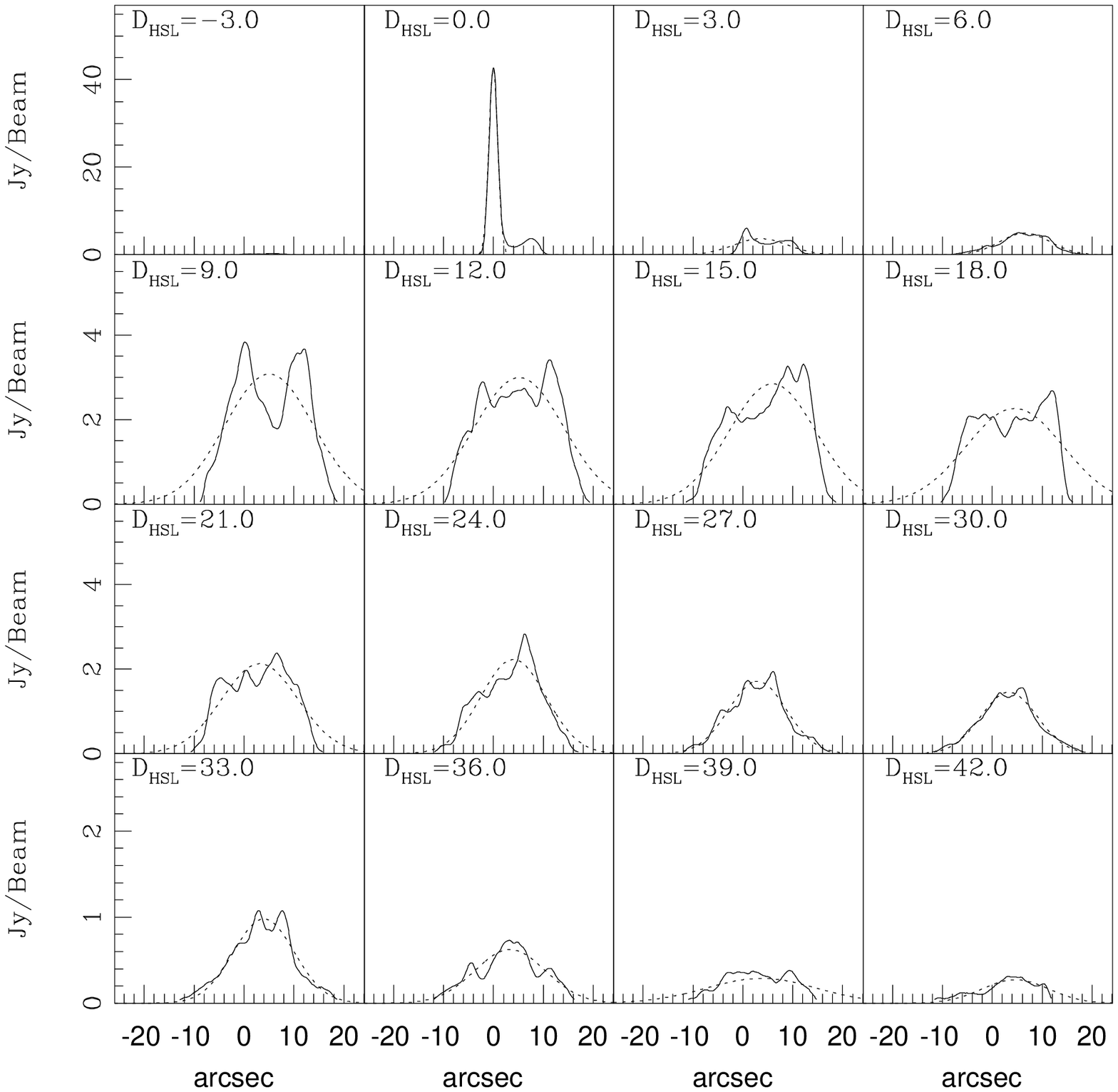}
\caption{Cross-sectional surface brightness profiles (solid lines)
and Gaussian fits (dashed lines)
for the Left side of Cygnus A at
151 MHz (left panel) and 1345 MHz (right panel).
The distance of the slice from the hot spot is indicated
for each slice.
\label{fig1L}}
\end{figure}

\clearpage
\begin{figure}
\plottwo{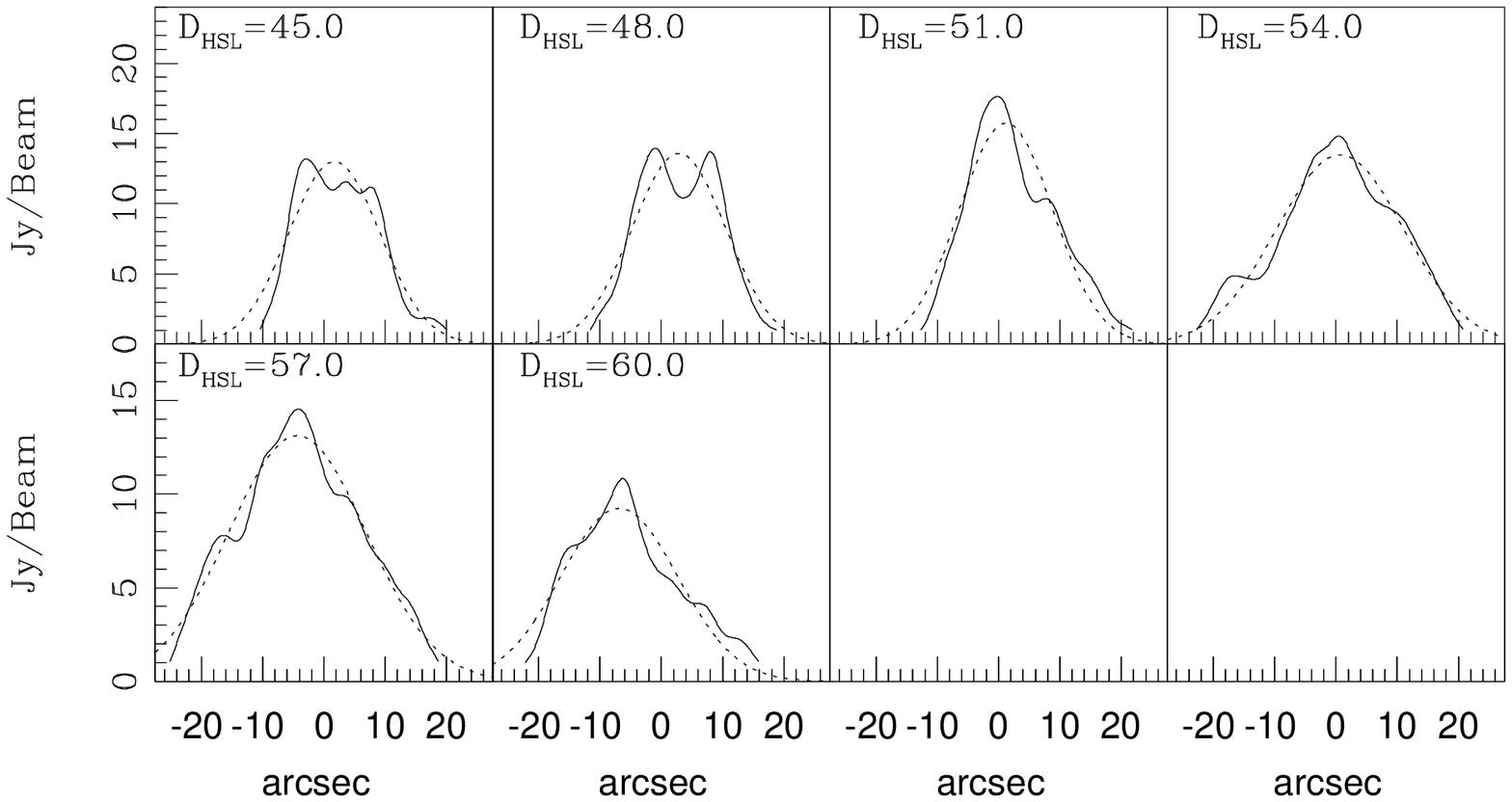}{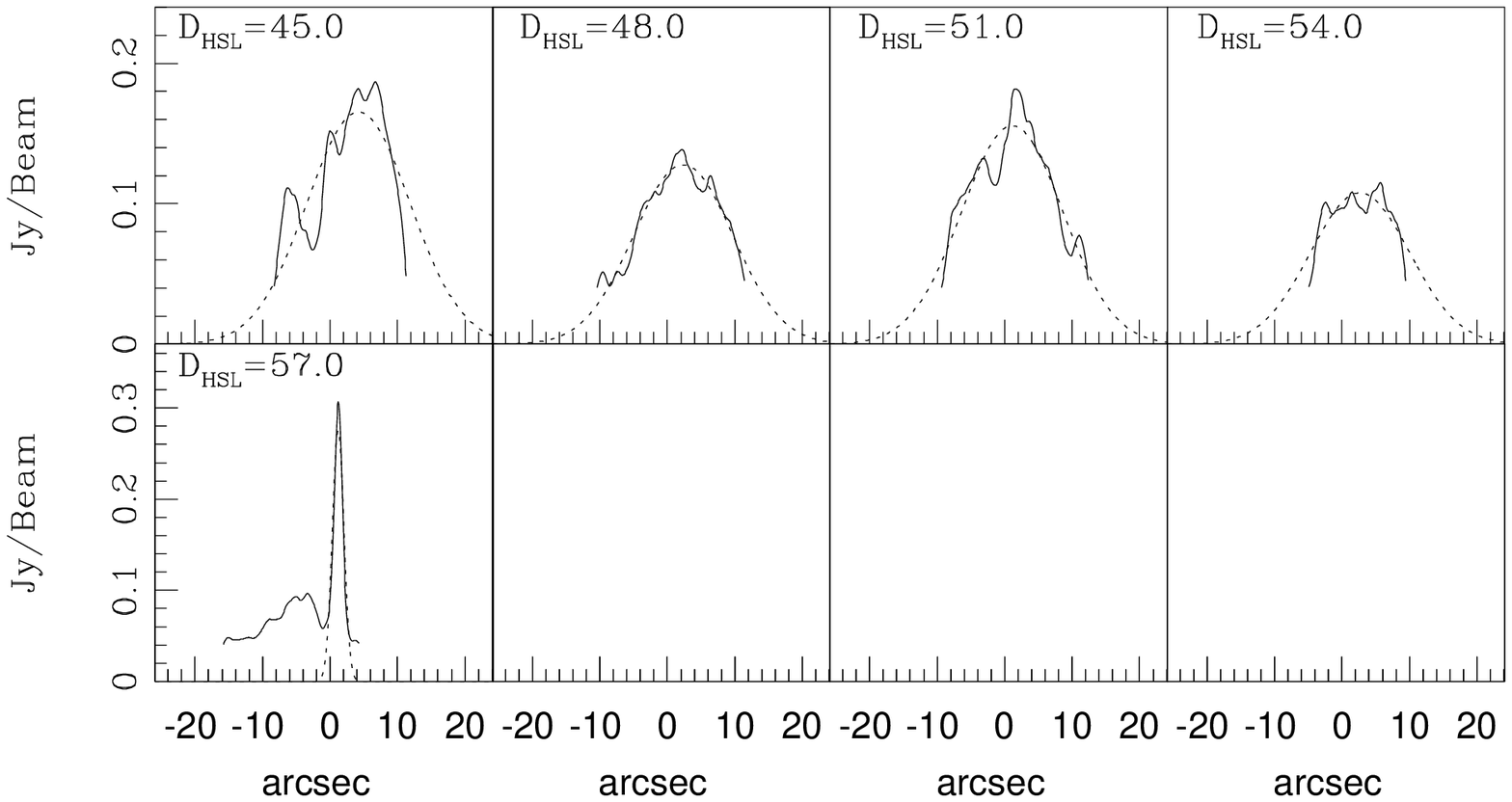}
\caption{Continuation of the cross-sectional surface
brightness profiles and Gaussian
fits for the Left side of Cygnus A at
151 MHz (left panel) and 1345 MHz (right panel).
\label{fig2L}}
\end{figure}

\clearpage

\begin{figure}
\plottwo{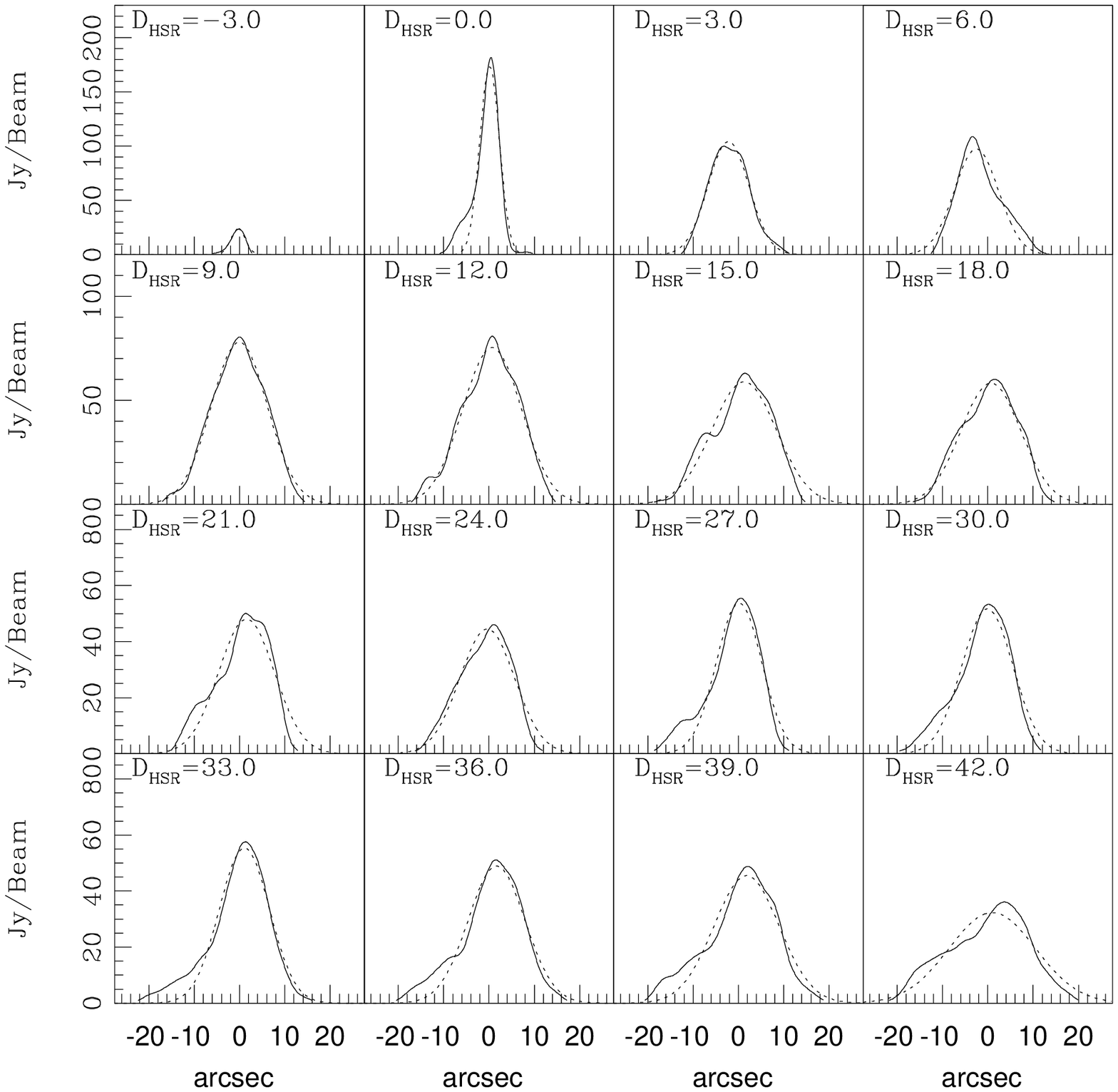}{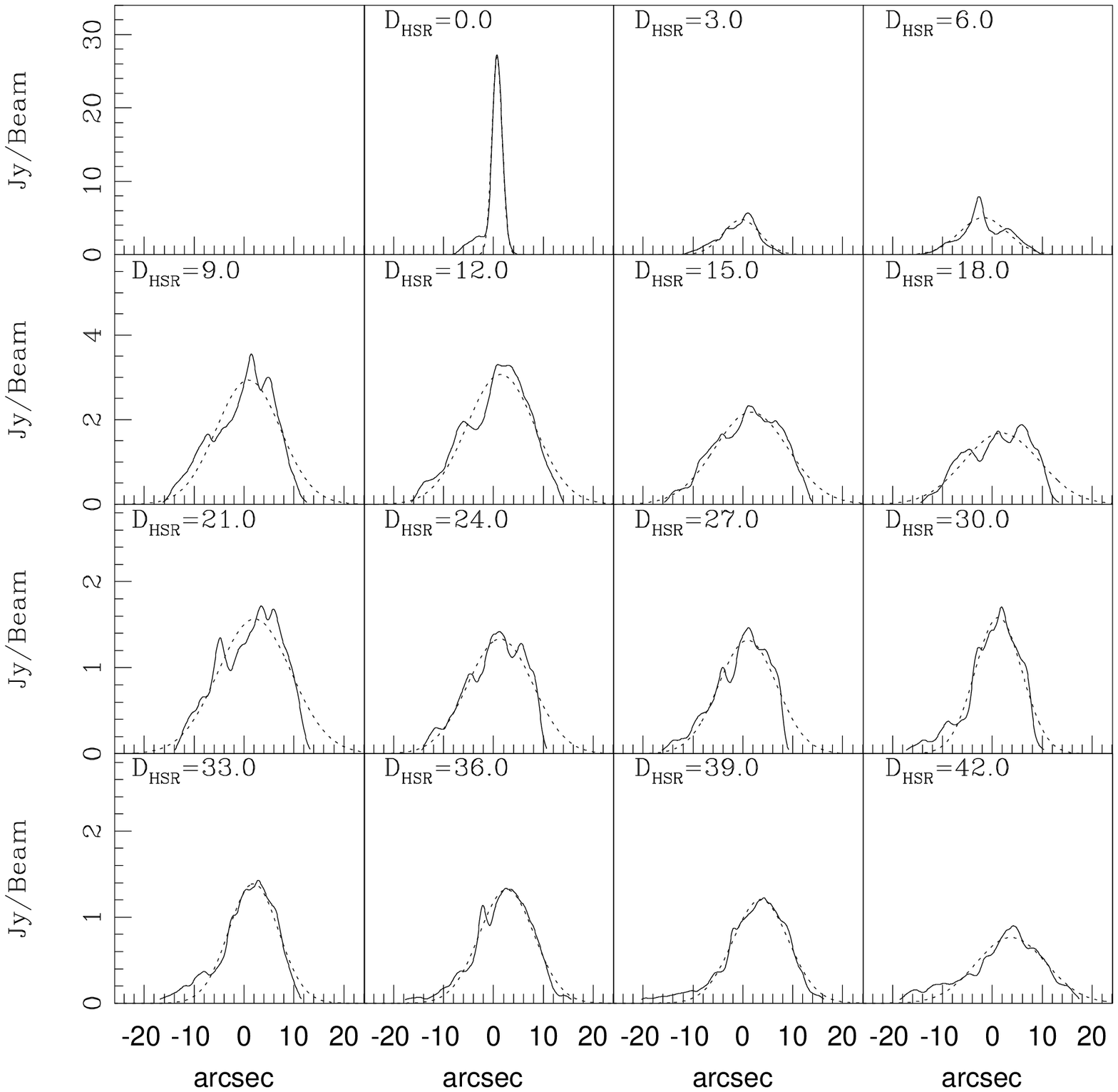}
\caption{Cross-sectional surface brightness profiles (solid lines)
and Gaussian fits (dashed lines)
for the Right side of Cygnus A at
151 MHz (left panel) and 1345 MHz (right panel).
The distance of the slice from the hot spot is indicated
for each slice.  \label{fig1R}}
\end{figure}

\clearpage
\begin{figure}
\plottwo{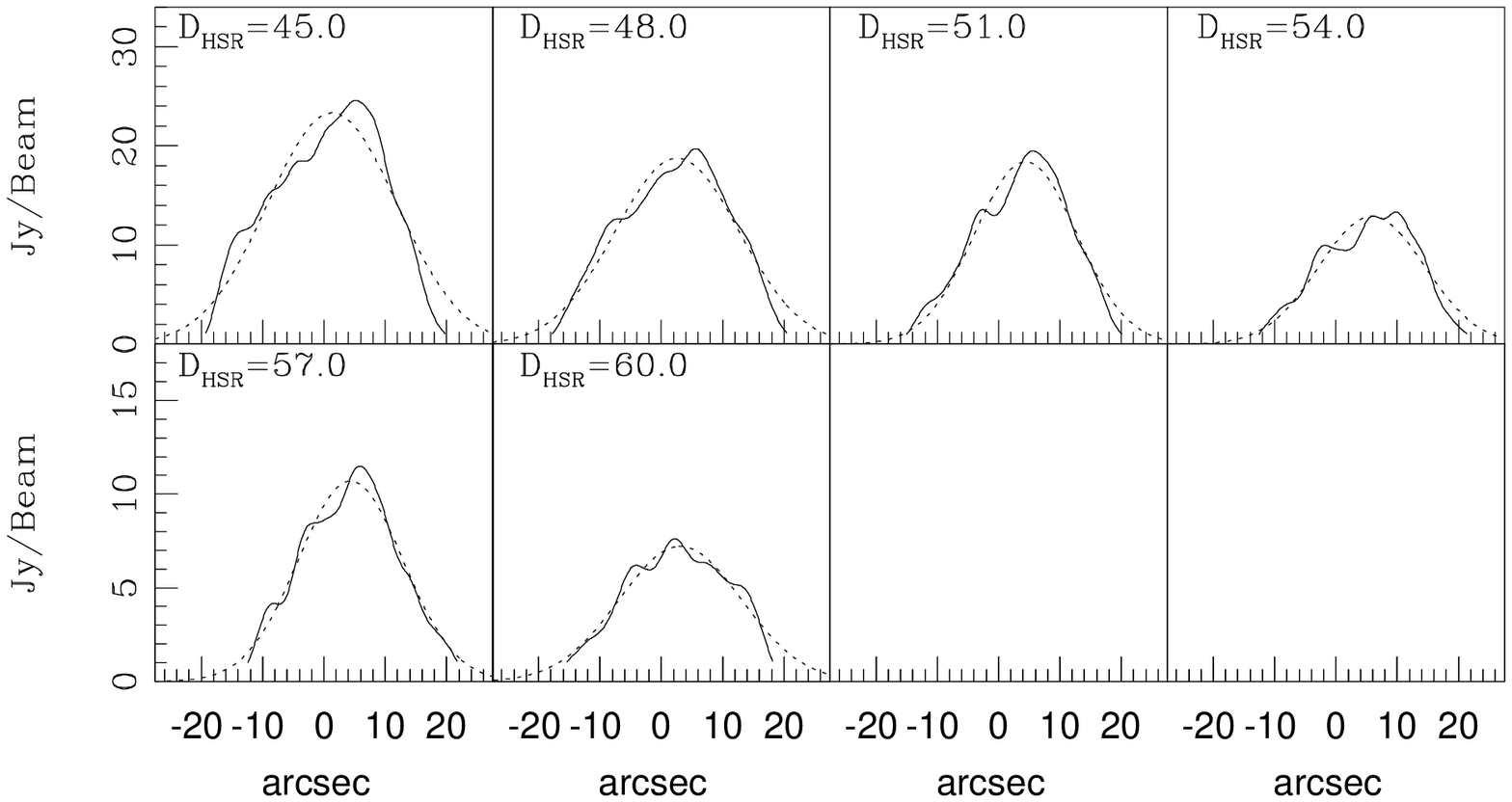}{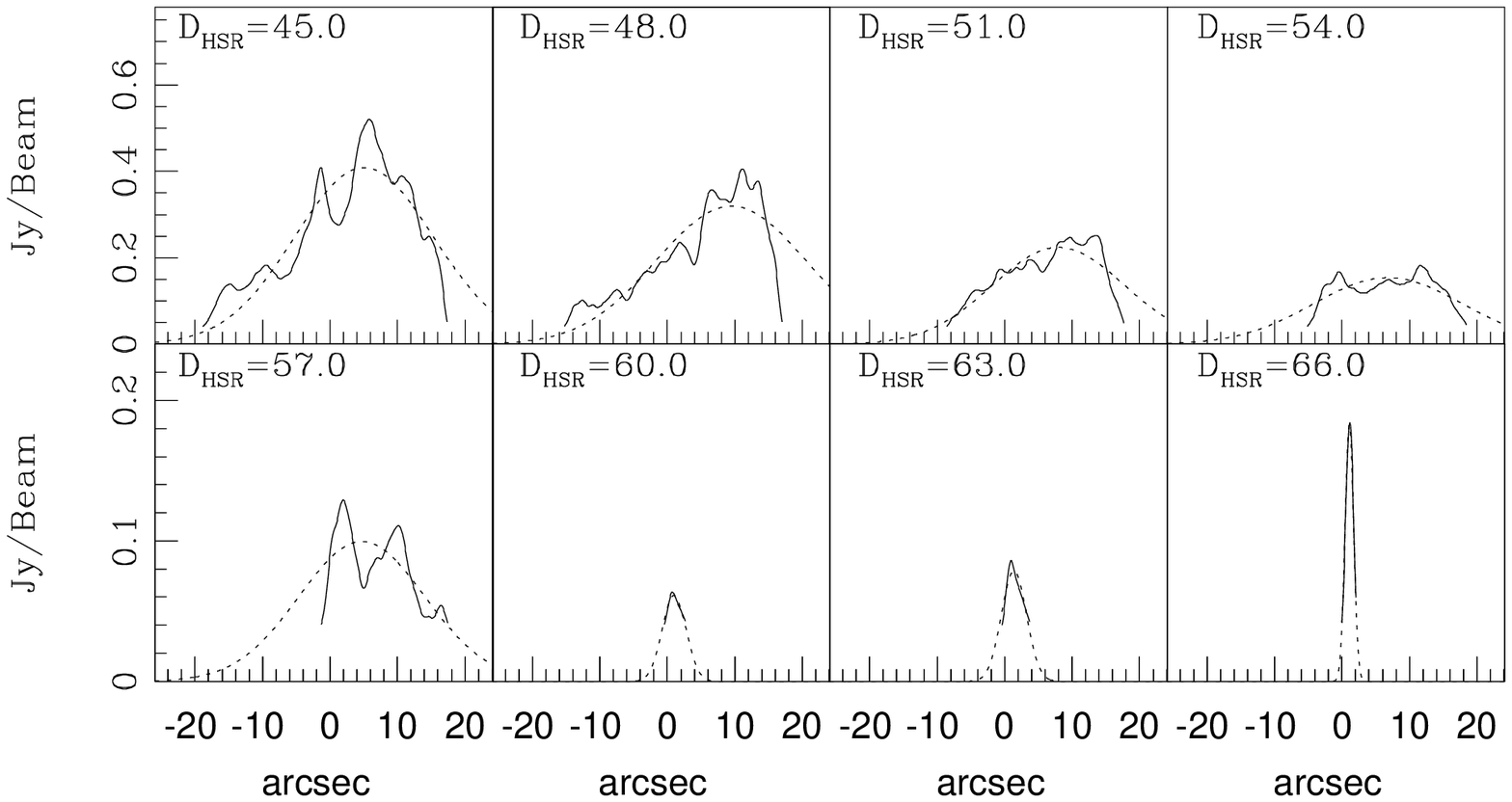}
\caption{Continuation of the cross-sectional surface
brightness profiles and Gaussian
fits  for the Right side of Cygnus A at
151 MHz (left panel) and 1345 MHz (right panel).
\label{fig2R}}
\end{figure}

\clearpage

\begin{figure}
\caption{The full width at half max of the best fit Gaussian
as a function of slice distance from the hot spot
at 151 MHz (left panel) and 1345 MHz (right panel).
The upper panels show results from the Left side of the
source, while the lower panels show results from the Right
side.
\label{fig3}}
\end{figure}

\clearpage
\begin{figure}
\caption{Ratio of the FWHM obtained at 151 MHz to that obtained
at 1345 MHz. A more narrowly focused view is shown on the right panel.
\label{fig4}}
\end{figure}

\clearpage

\begin{figure}
\caption{The first moment of the cross-sectional surface brightness
profiles as a function of distance from the hot spot.
The 151 MHz data is shown on the left panels, while the 1345 MHz
data is shown on the right panels; the Left
side of the source is shown in the upper panels, and the Right side
is shown in the lower panels.
\label{fig5}}
\end{figure}

\clearpage
\begin{figure}
\caption{The second moment of the cross-sectional surface brightness
profiles as a function of distance from the hot spot for the
151 MHz data (left panel) and the 1345 MHz data (right panel).
The upper panels show results from the Left side of the source
and the lower panels show results from the Right side of the source.
\label{fig6}}
\end{figure}

\clearpage

\begin{figure}
\caption{The ratio of the Gaussian FWHM to the second moment
as a function of distance from the hotspot for the
151 MHz data (left panel) and the 1345 MHz data (right panel).
Measurements from the left side of the source are shown in the upper
panels; the average values of these are $2.59 \pm 0.07$ and
$2.45 \pm 0.03$ for the 151 and 1345 MHz data respectively.
Measurements from the right side are shown in the lower panels;
the average values are $2.49 \pm 0.07$ and $2.53 \pm 0.03$
for the 151 and 1345 MHz data respectively.
\label{width-ratio}}
\end{figure}

\clearpage \begin{figure}
\caption{The average surface brightness in a cross-sectional
surface brightness slice as a function of distance from the hot
spot for the 151 MHz data (left panel) and the 1345 MHz data (right panel)
obtained assuming the diameter of the bridge at each distance from the
hot spot can be approximated by the FWHM of the best fit Gaussian.
Results from the Left side of the source are shown in the upper panels,
while those from the Right are shown in the lower panels.
\label{fig8}} \end{figure}

\clearpage \begin{figure}
\caption{The average pressure in a cross-sectional
surface brightness slice as a function of distance from the hot
spot for the 151 MHz data (left) and the 1345 MHz data (right).
The pressure for the 151 MHz data has units of
$1.2 \times 10^{-11} \hbox{ erg cm}^{-3}(1.33b^{-1.5}+b^2)$,
where $b$ parameterizes the offset from minimum energy conditions.
For minimum energy conditions, $b=1$, and the normalization
factor or pressure unit
for the 151 MHz data is $2.8 \times 10^{-11} \hbox{ erg cm}^{-3}$,
while for a field strength that is about 0.25 the minimum energy
value, the normalization is $1.3 \times 10^{-10} \hbox{ erg cm}^{-3}$.
The normalization factor for the 1345 MHz data is
$10^{-10} \hbox{ erg cm}^{-3}(1.33b^{-1.5}+b^2)$.  For
minimum energy conditions the pressure unit for the
1345 MHz data is $2.4 \times 10^{-10} \hbox{ erg cm}^{-3}$,
while for $b=0.25$, the pressure unit is
$1.1 \times 10^{-9} \hbox{ erg cm}^{-3}$.
Measurements from the Left side of the source are shown in the
upper panels, while those from the Right side are shown in the lower panels.
\label{fig9}} \end{figure}

\clearpage
\begin{figure}
\caption{The average minimum energy
magnetic field strength in a cross-sectional
surface brightness slice as a function of distance from the hot
spot for the 151 MHz data (left) and the 1345 MHz data (right).
The units of the minimum magnetic field strength for the
151 MHz data are 30 $\mu$G, while those for the 1345 MHz data
are 88 $\mu$G.  Estimates from the Left side of the source
are shown in the upper panels, while those from the Right side
of the source are shown in the lower panels.
\label{fig10}}
\end{figure}

\clearpage
\begin{figure}
\caption{Method of obtaining the emissivity as a function of
radial distance from the slice center.
\label{fig13}}
\end{figure}

\clearpage

\begin{figure}
\plottwo{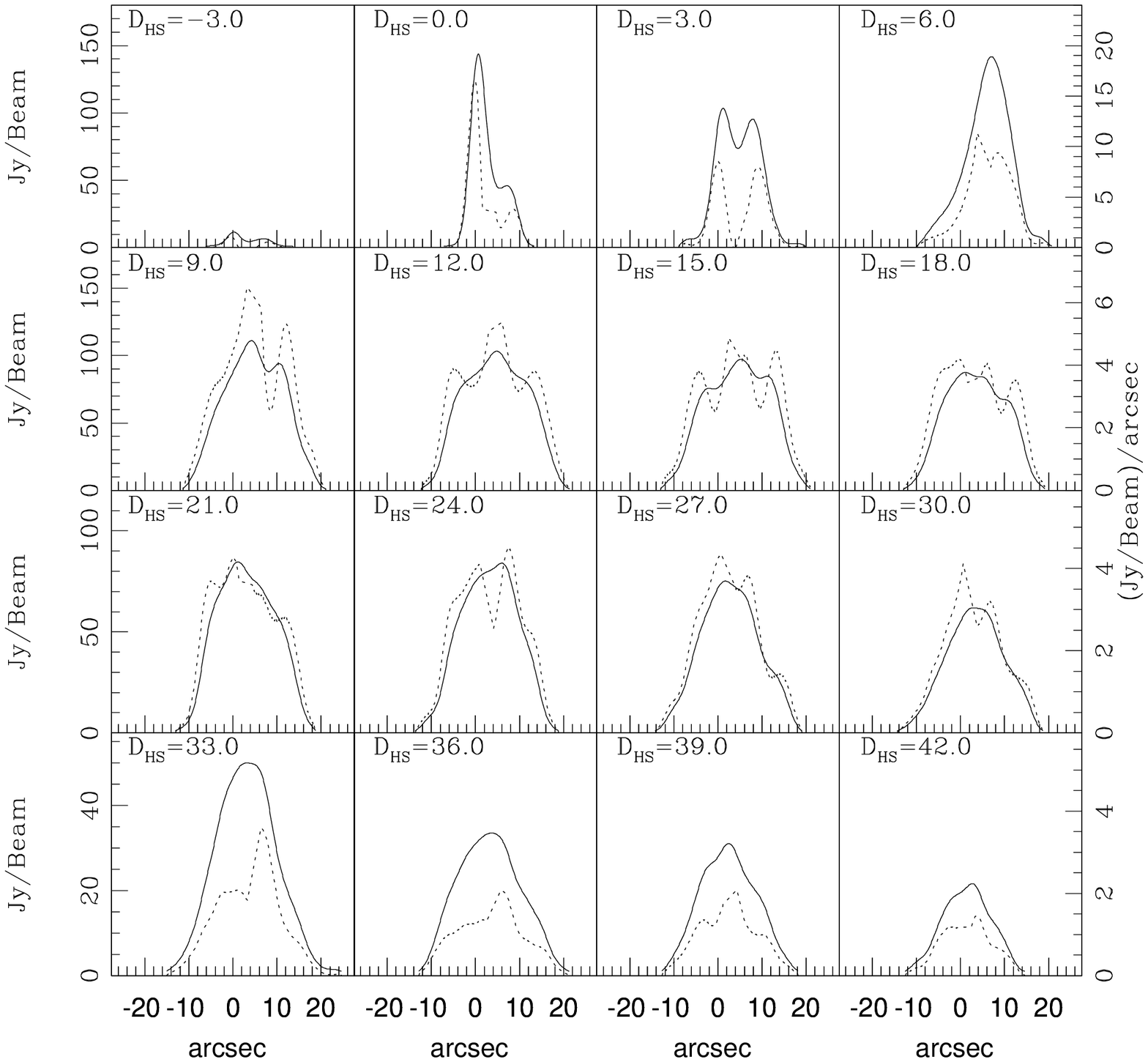}{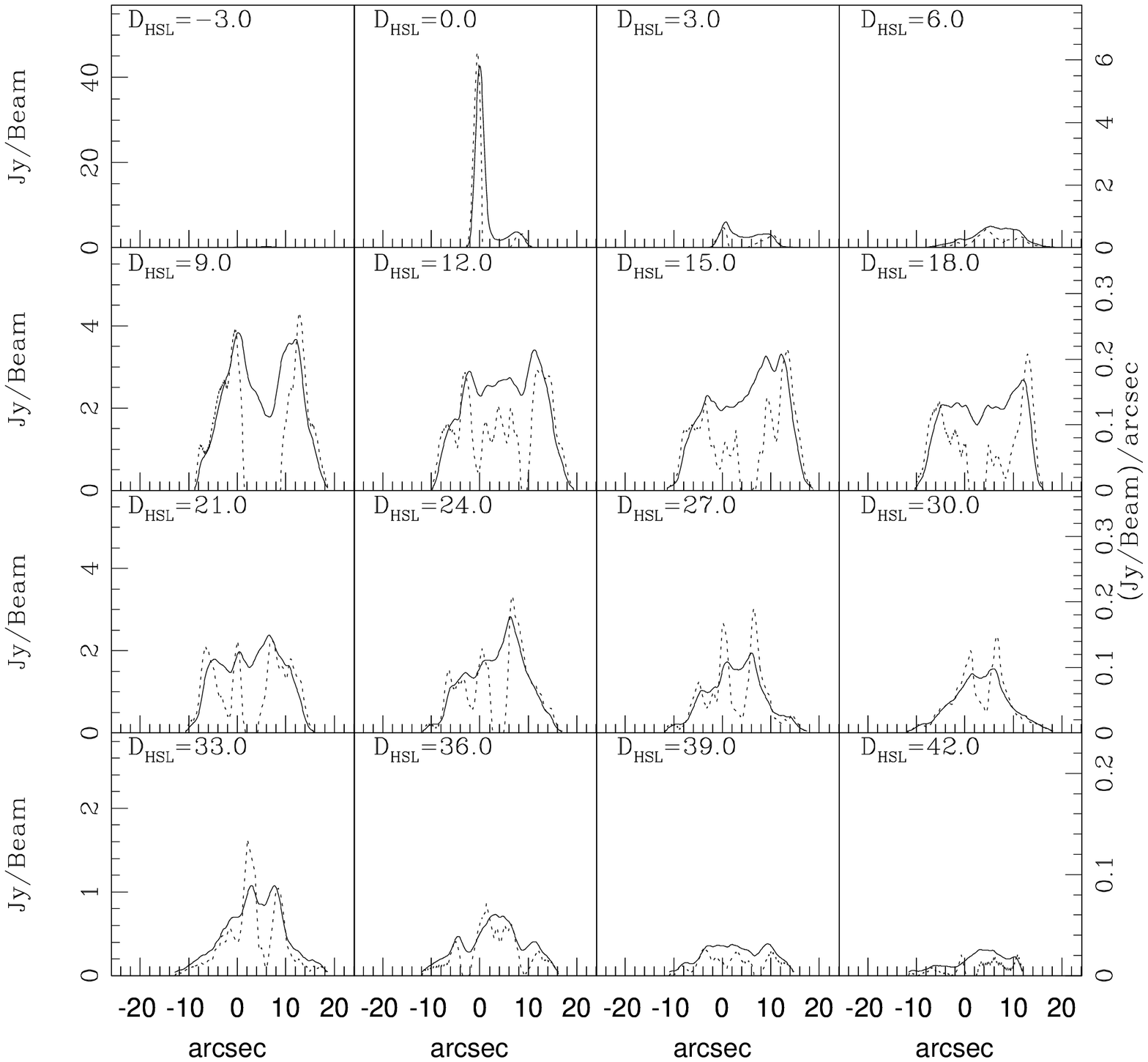}
\caption{The radio emissivities (dashed lines) and
cross-sectional surface brightness profiles (solid lines)
for the 151 MHz data (left panel) and the 1345 MHz data (right panel)
for the Left side of Cygnus A.
The units for the 151 MHz data are $1.8 \times 10^{-34} \hbox{ erg s}^{-1}
\hbox{ cm}^{-3} \hbox{ Hz}^{-1}$, while those for the 1345 MHz
data are  $9.0 \times 10^{-34} \hbox{ erg s}^{-1}
\hbox{ cm}^{-3} \hbox{ Hz}^{-1}$.
\label{figSBEML1}}
\end{figure}

\clearpage
\begin{figure}
\plottwo{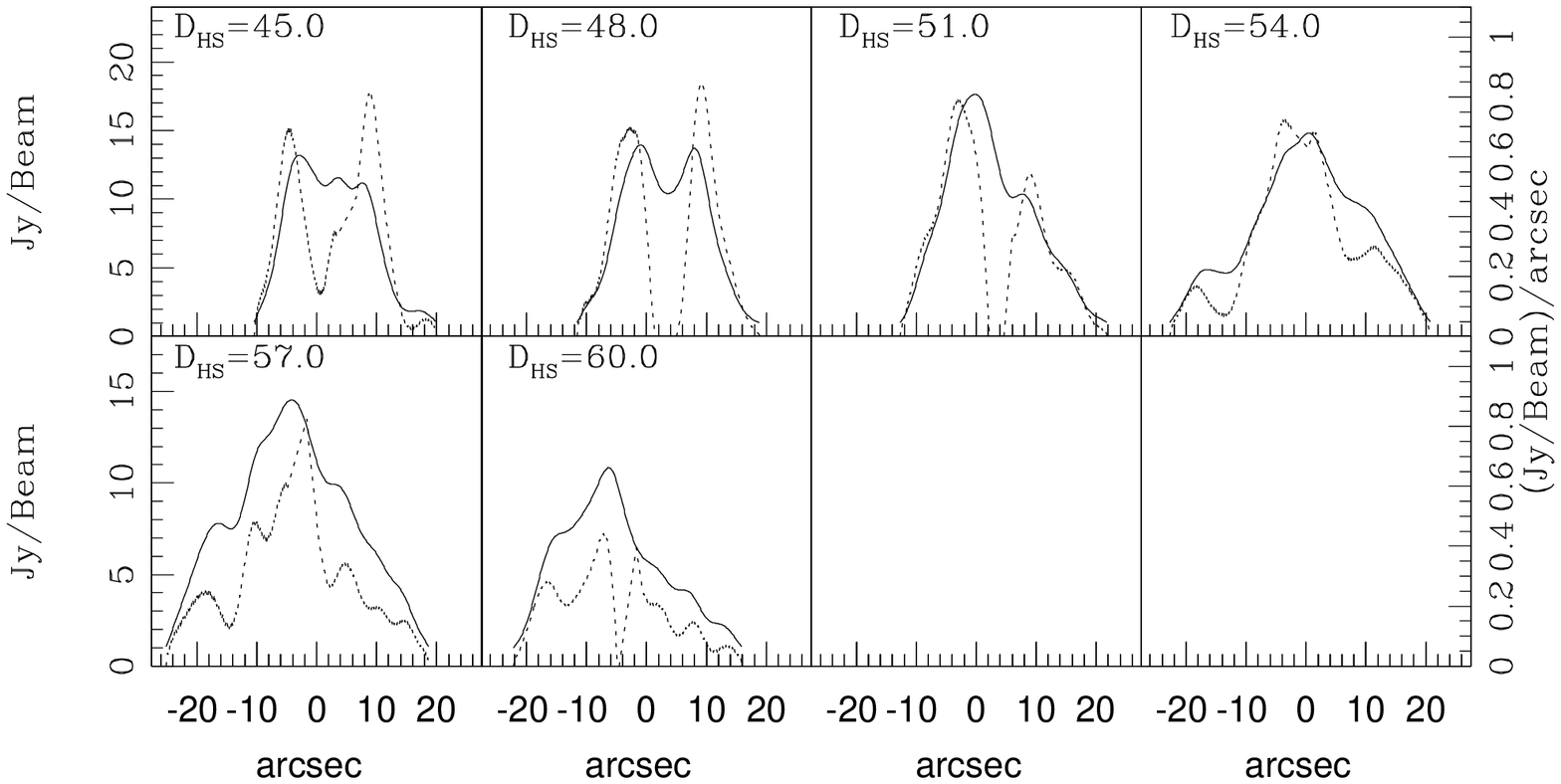}{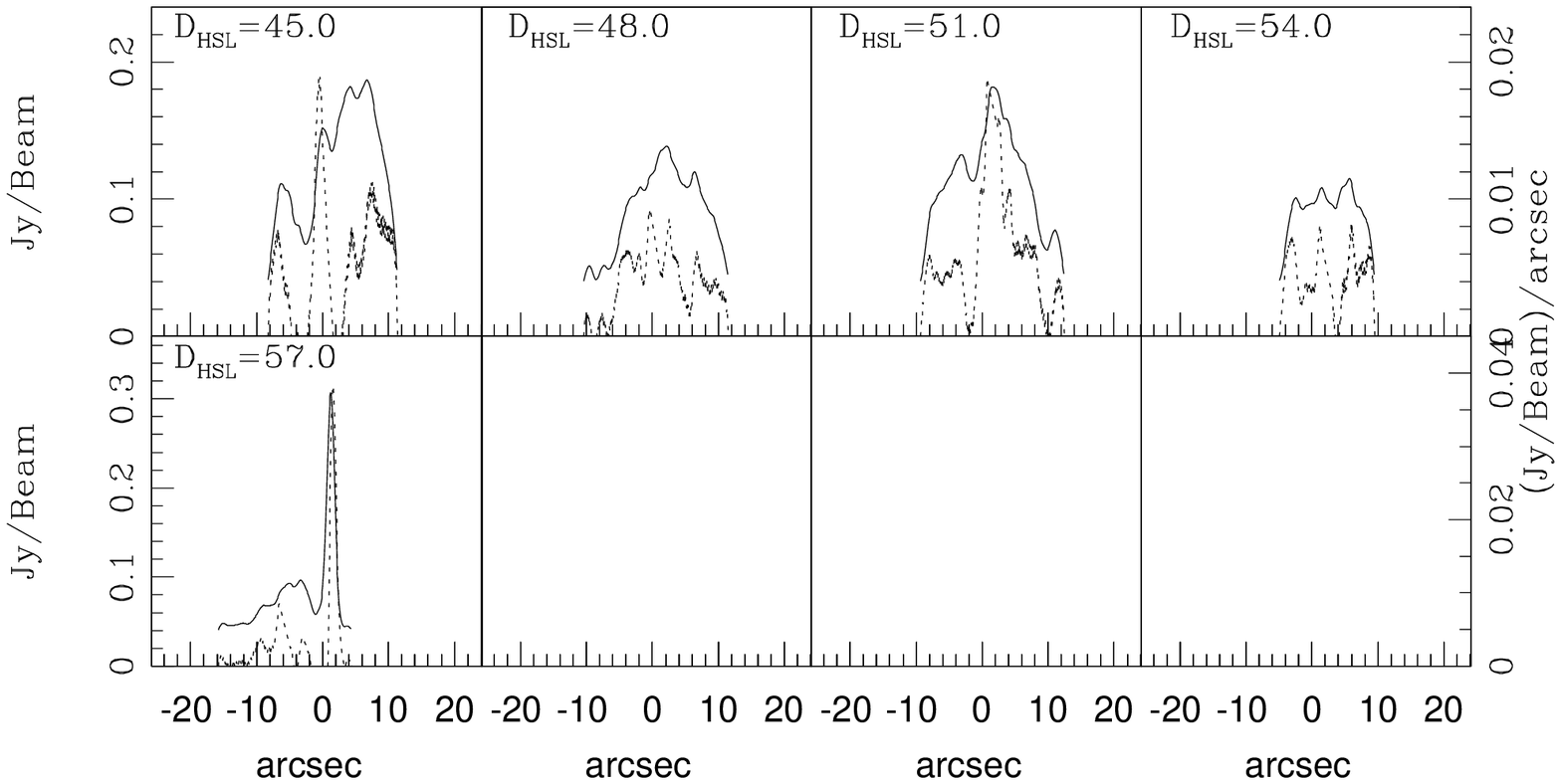}
\caption{Continuation of the radio emissivities
and cross-sectional surface
brightness profiles for the Left side of
Cygnus A.\label{figSBEML2}}
\end{figure}

\clearpage

\begin{figure}
\plottwo{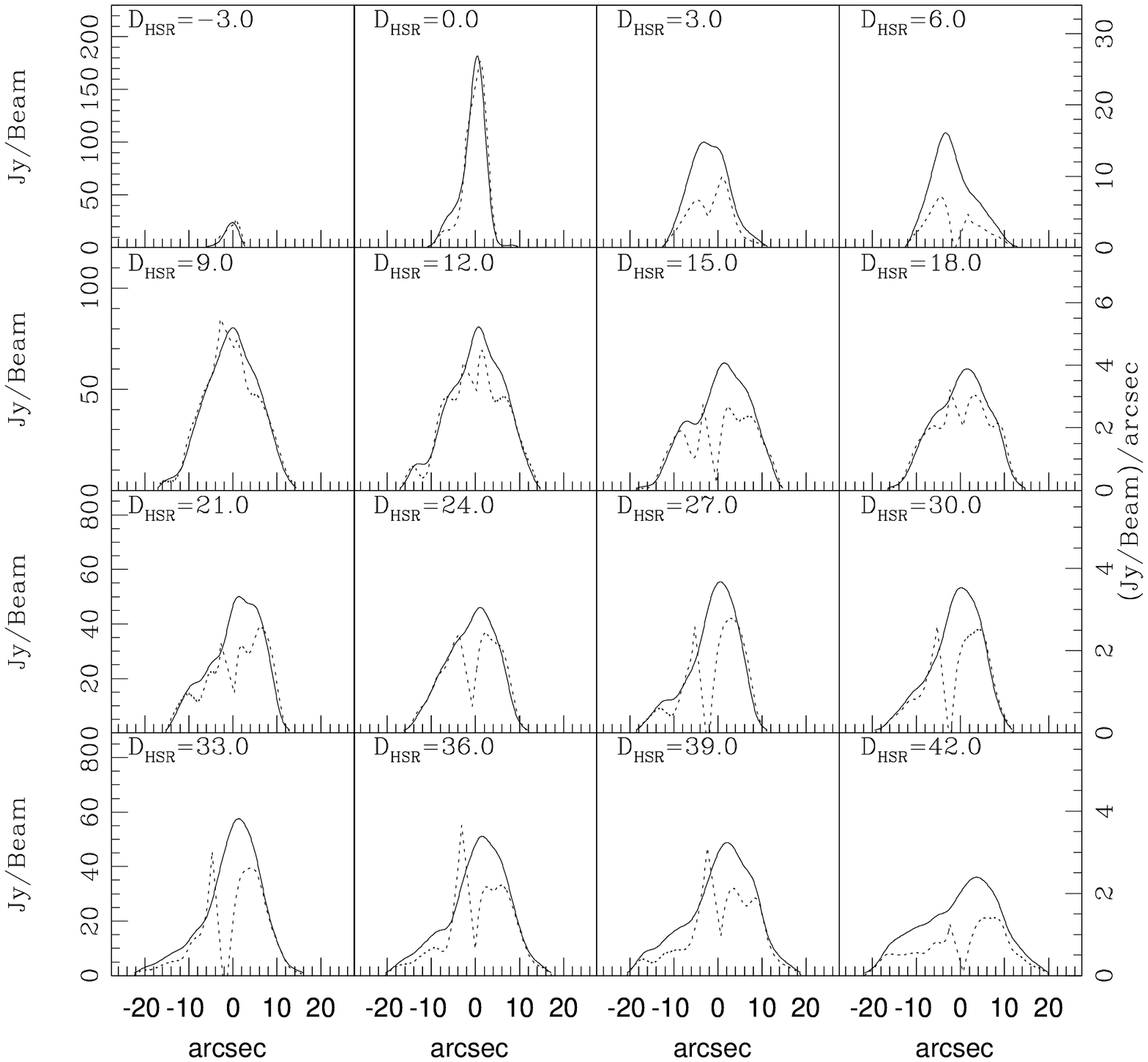}{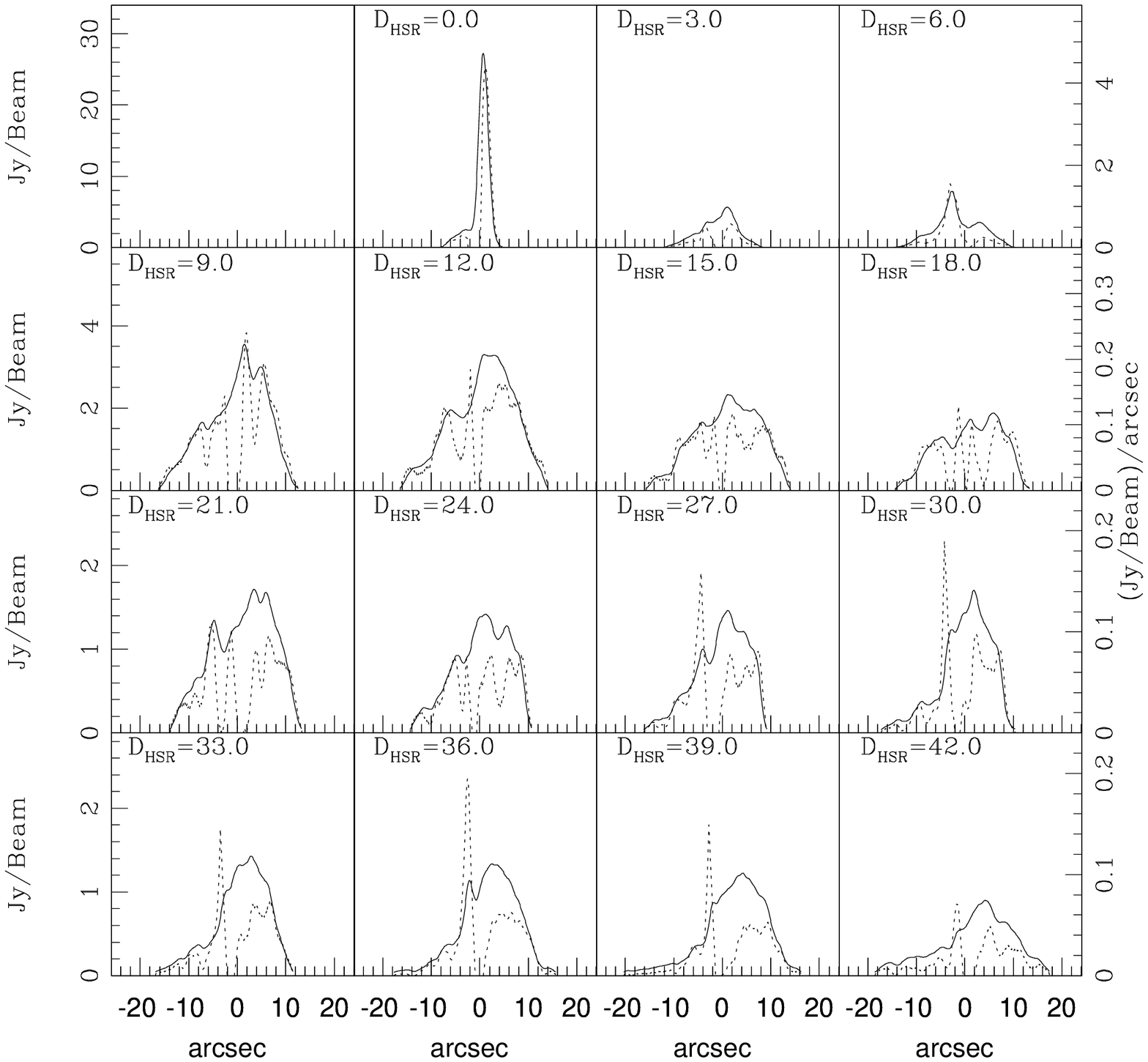}
\caption{The radio emissivities (dashed lines) and
cross-sectional surface brightness profiles (solid lines)
for the 151 MHz data (left panel) and the 1345 MHz data (right panel)
for the Right side of Cygnus A.
The units for the 151 MHz data are $1.8 \times 10^{-34} \hbox{ erg s}^{-1}
\hbox{ cm}^{-3} \hbox{ Hz}^{-1}$, while those for the 1345 MHz
data are  $9.0 \times 10^{-34} \hbox{ erg s}^{-1}
\hbox{ cm}^{-3} \hbox{ Hz}^{-1}$.
\label{figSBEMR1}}
\end{figure}

\clearpage
\begin{figure}
\plottwo{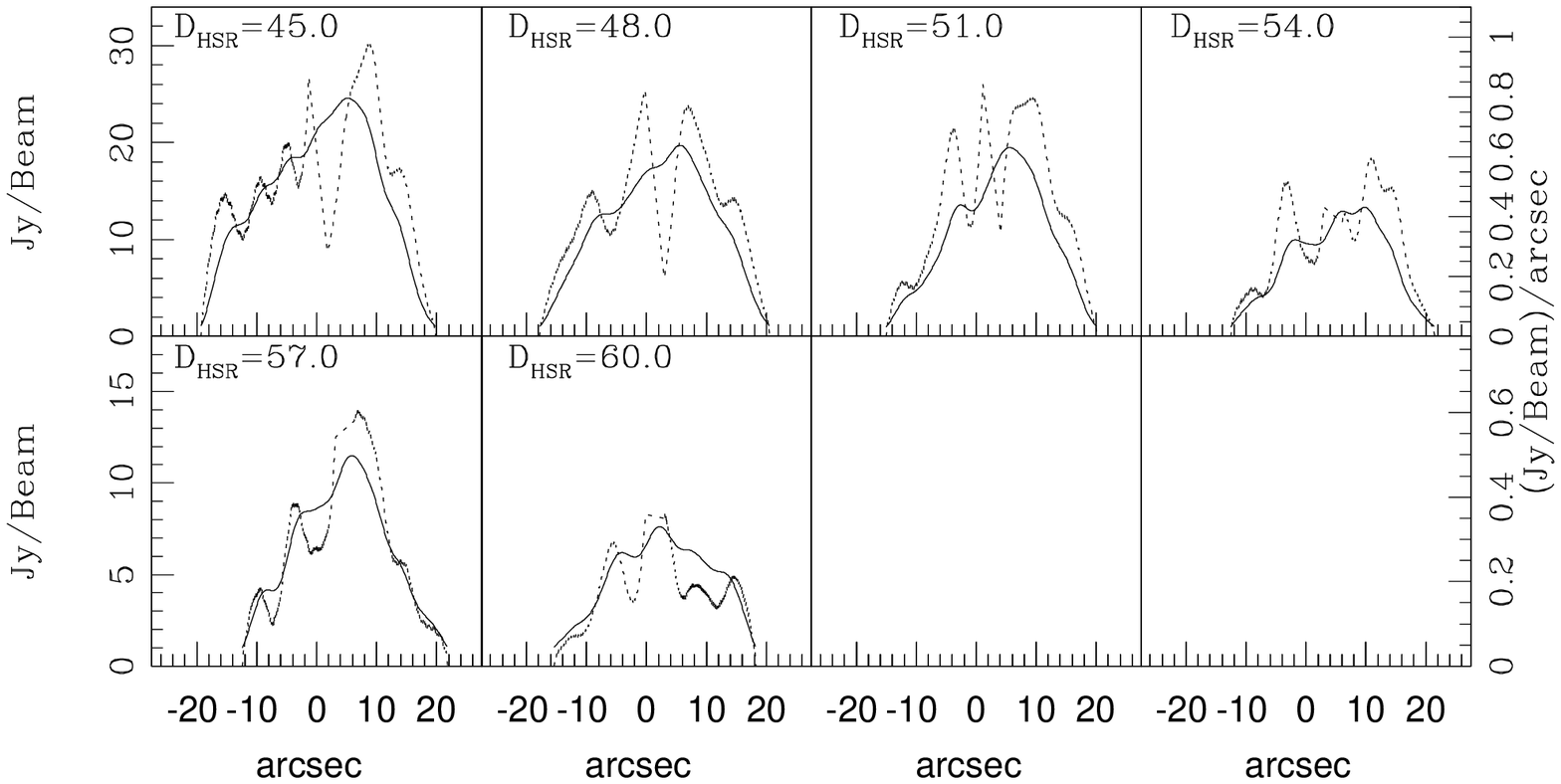}{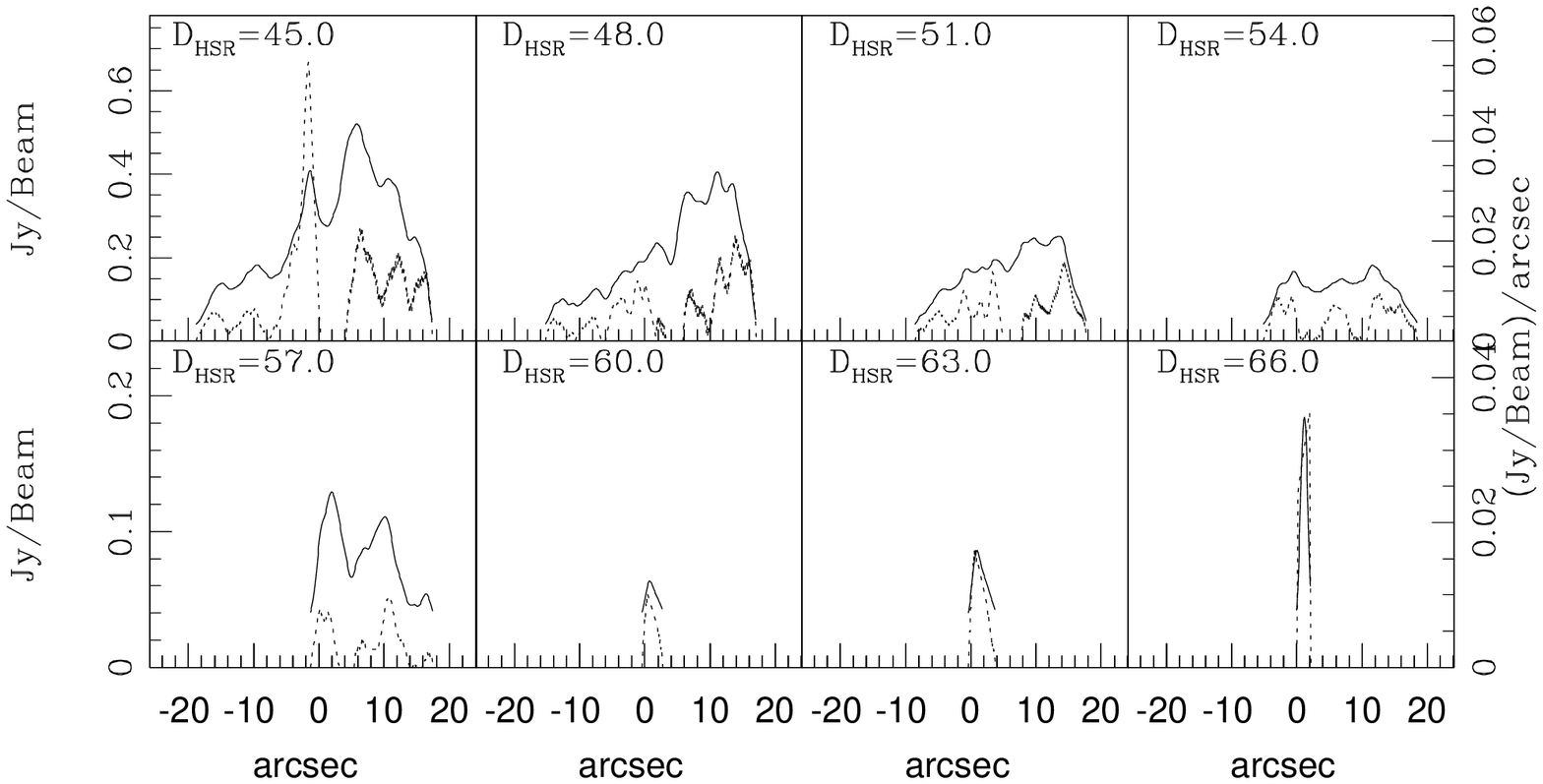}
\caption{Continuation of the and radio emissivities
and cross-sectional surface
brightness profiles and radio emissivities
for the Right side of Cygnus A.\label{figSBEMR2}}
\end{figure}
\clearpage

\begin{figure}[ht]
\epsscale{0.6}
\figcaption{Density contour map of a jet  propagating in a
constant density atmosphere for Run D01 ($M=10, \eta=2\times
10^{-4}$). The grid size
is 66 $\times 79$ kpc.
\label{ASUD_IndvPnlD01} }
\end{figure}
\clearpage

\begin{figure}[ht]
\epsscale{0.8}
\plotone{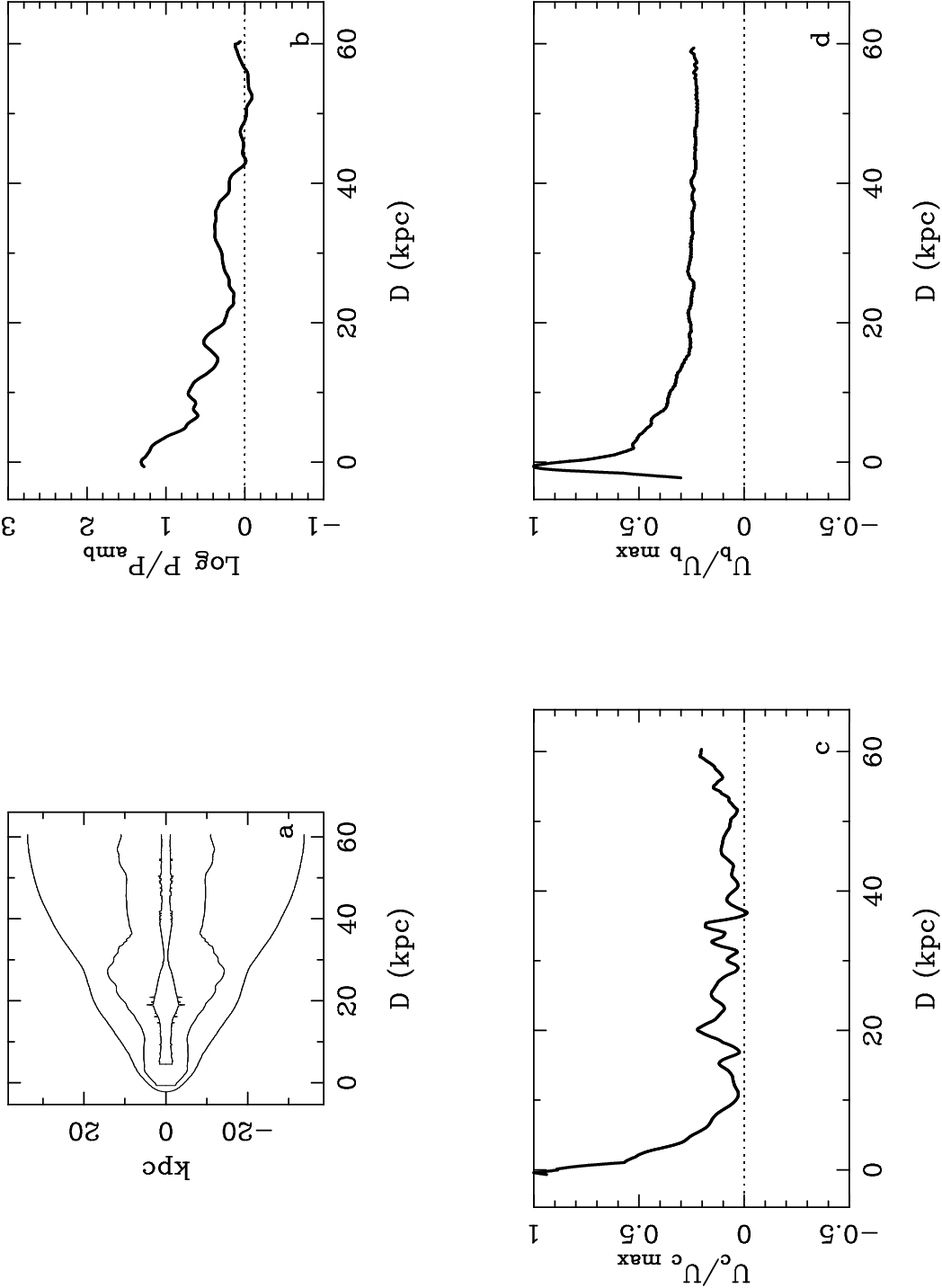}
\figcaption{(a) Contour of the bow shock, cocoon and jet for the simulation
shown in Fig. \ref{ASUD_IndvPnlD01} (Run D01). Panel (b) shows the longitudinal
profile of
the average cocoon pressure, cocoon lateral speed (c) and bow shock lateral
speed (d).
 \label{ACygA-ViewD01} }
\end{figure}
\clearpage

\begin{figure}[ht]
\plotone{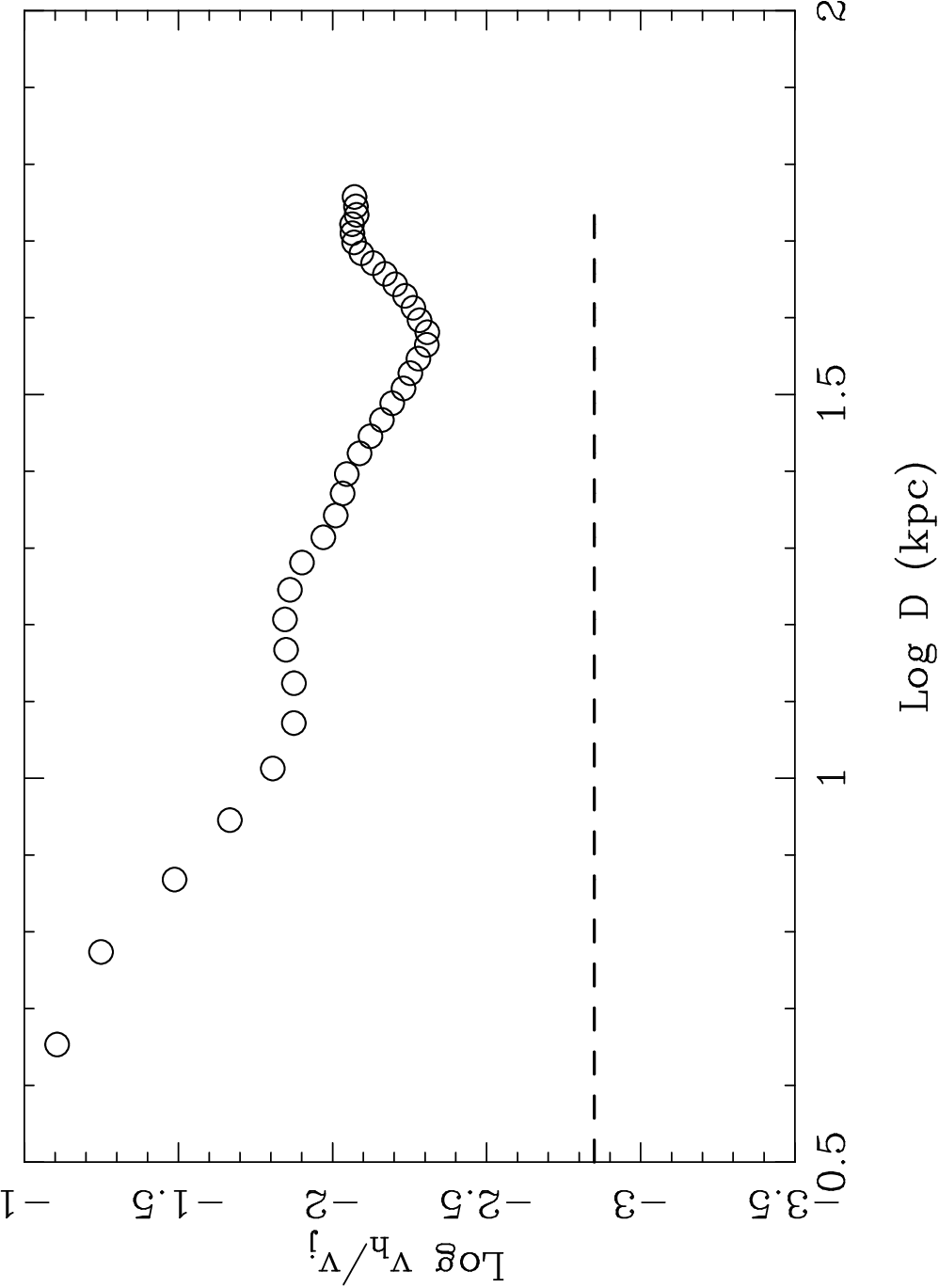} \epsscale{0.8}
\figcaption{Advance speed of the head (open circles) relative to the 
jet initial speed as a function of source size. The dashed line 
represents the ambient sound speed.
\label{AHeadSpeedD01} }
\end{figure}
\clearpage

\begin{figure}[ht]
\figcaption{Synthetic radio image showing the time evolution of the
numerical simulation of a jet ($M=10, \eta=2\times 10^{-4}$) propagating
in a constant density atmosphere. Each panel measures 144 $\times
54$ kpc and the synthetic data have been smoothed by a Gaussian with 
FWHM of 3 arcsec to match the 151 MHz data, as shown by the black circle 
in panel e.  The age of the source corresponding to each panel is: 
(a) 1.35 Myr, (b) 4.53 Myr, (c) 6.17 Myr, (d) 8.89 Myr and (e) 9.15 Myr.
One can clearly see the
characteristic edge-brightened shape of powerful FR-II sources and
an aspect ratio resembling that of Cygnus A.
 \label{AradioD01te} }
\end{figure}
\clearpage

\begin{figure}[ht]
\epsscale{0.8} \plotone{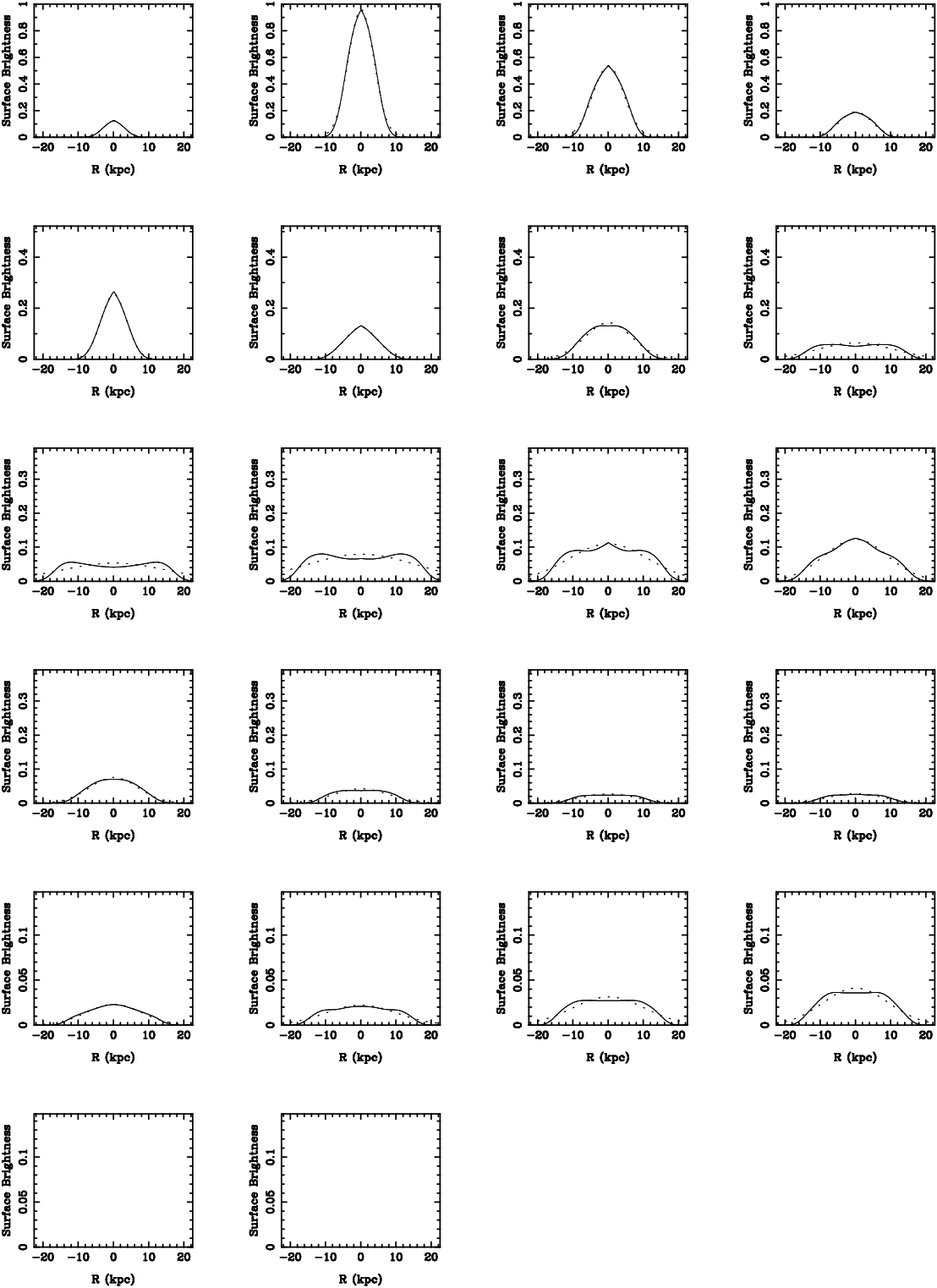} \figcaption{
Cross-sectional surface brightness profiles (solid lines) and
Gaussian fits (dotted lines) for the numerical simulation (Run
D01). The distance between each panel is 3 kpc.
 \label{ACygA-SBD01te} }
\end{figure}
\clearpage

\begin{figure}[ht]
\epsscale{0.8} \plotone{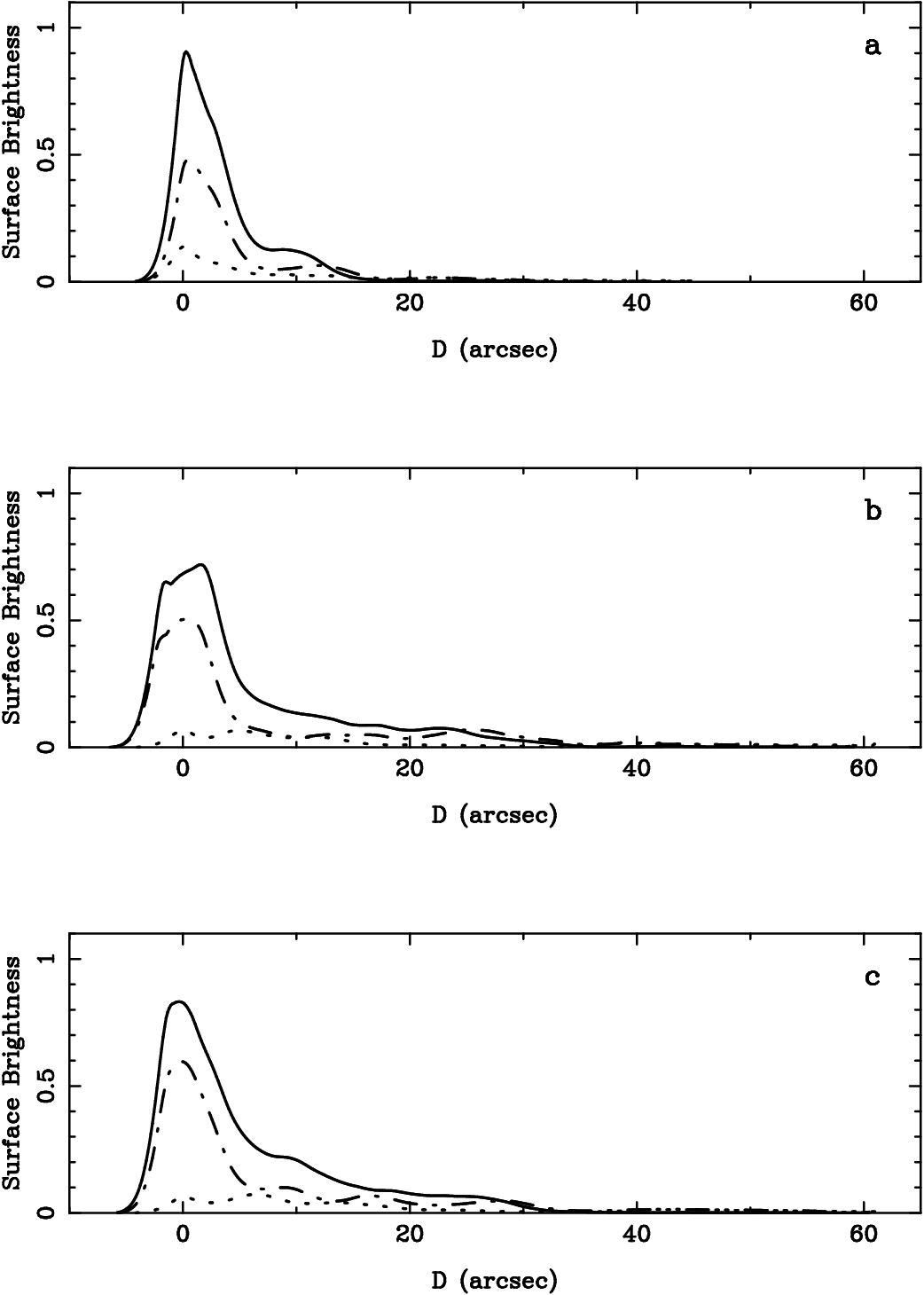}
\figcaption{ Average surface brightness at 151 MHz in a
cross-sectional slice
 as a function of distance
from the hot spot taking into account radiative losses. The radio
emissivity is calculated supposing the relativistic electron
density is proportional to the total pressure (full line), thermal
pressure (dashed-dotted line) and the kinetic pressure (dotted
line). Panels (a), (b) and (c) correspond to panels (c), (d) and
(e) of Fig. \ref{AradioD01te}, respectively.
\label{ACyg_S_toteki} }
\end{figure}
\clearpage

\begin{figure}[ht]
\epsscale{0.8} \plotone{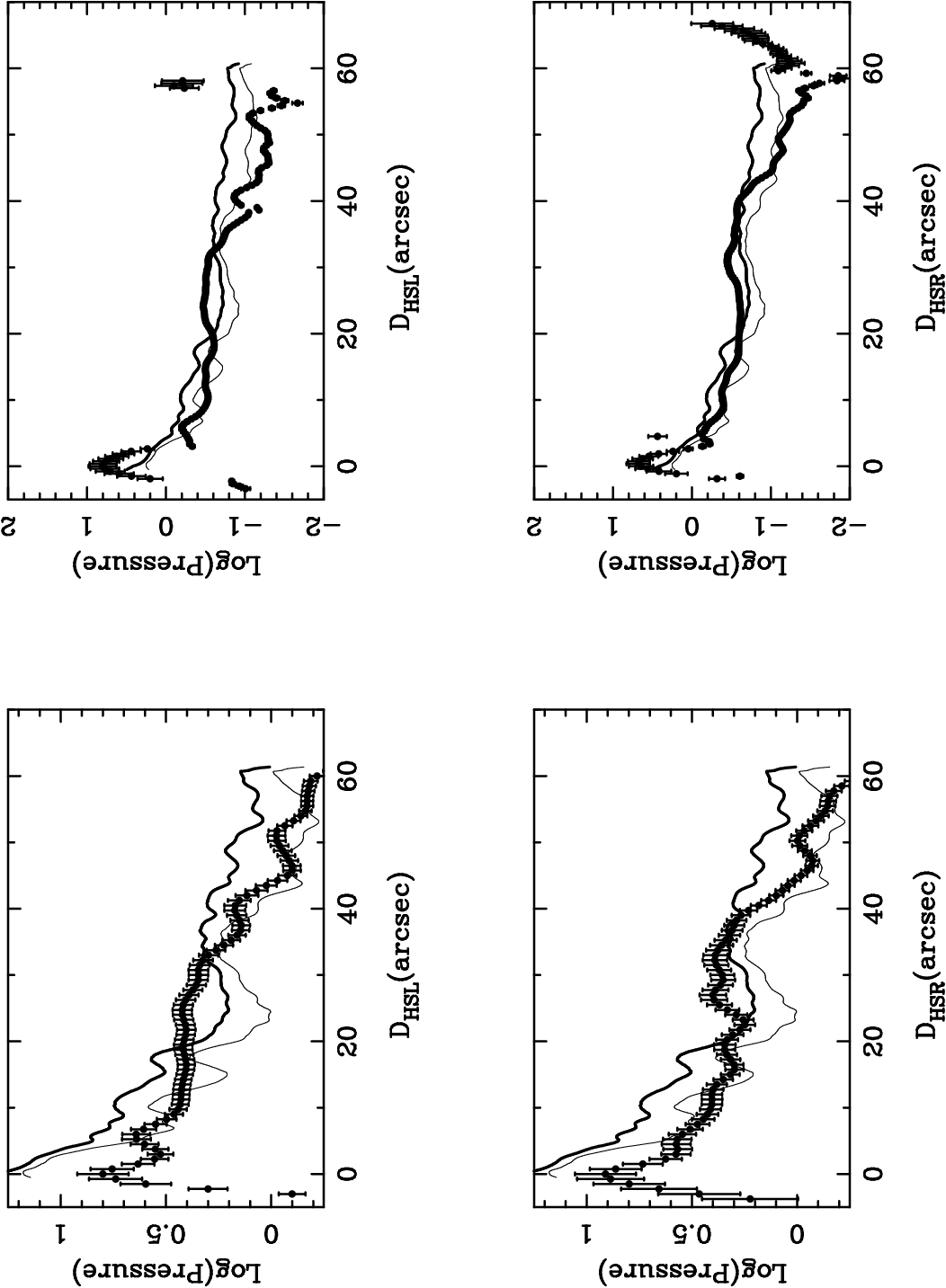}
\figcaption{ Comparison between the logarithm of the observed and
calculated average total (thermal plus kinetic - thick line) and
thermal (thin line) pressures as a function of distance from the
hot spot for 151 MHz (left panels) and 1345 MHz (right panels).
The Left side of the source is shown in the upper panels, and the
Right side is shown in the lower panels. The calculated pressure
is for an ambient density $n_a=0.01$ cm$^{-3}$, and the
units on the pressure are the same as in figure \ref{fig9} for 
a value of the parameter
$b$, defining the offset from minimum energy conditions, of 0.25.
Thus, the unit on the 151 MHz pressure is  
$1.3 \times 10^{-10} \hbox{ erg cm}^{-3}$, while that
on the 1345 MHz pressure is $1.1 \times 10^{-9} \hbox{ erg cm}^{-3}$.
 \label{ACyg_PressureAll} }
\end{figure}
\clearpage

\begin{figure}[ht]
\epsscale{0.8} \plotone{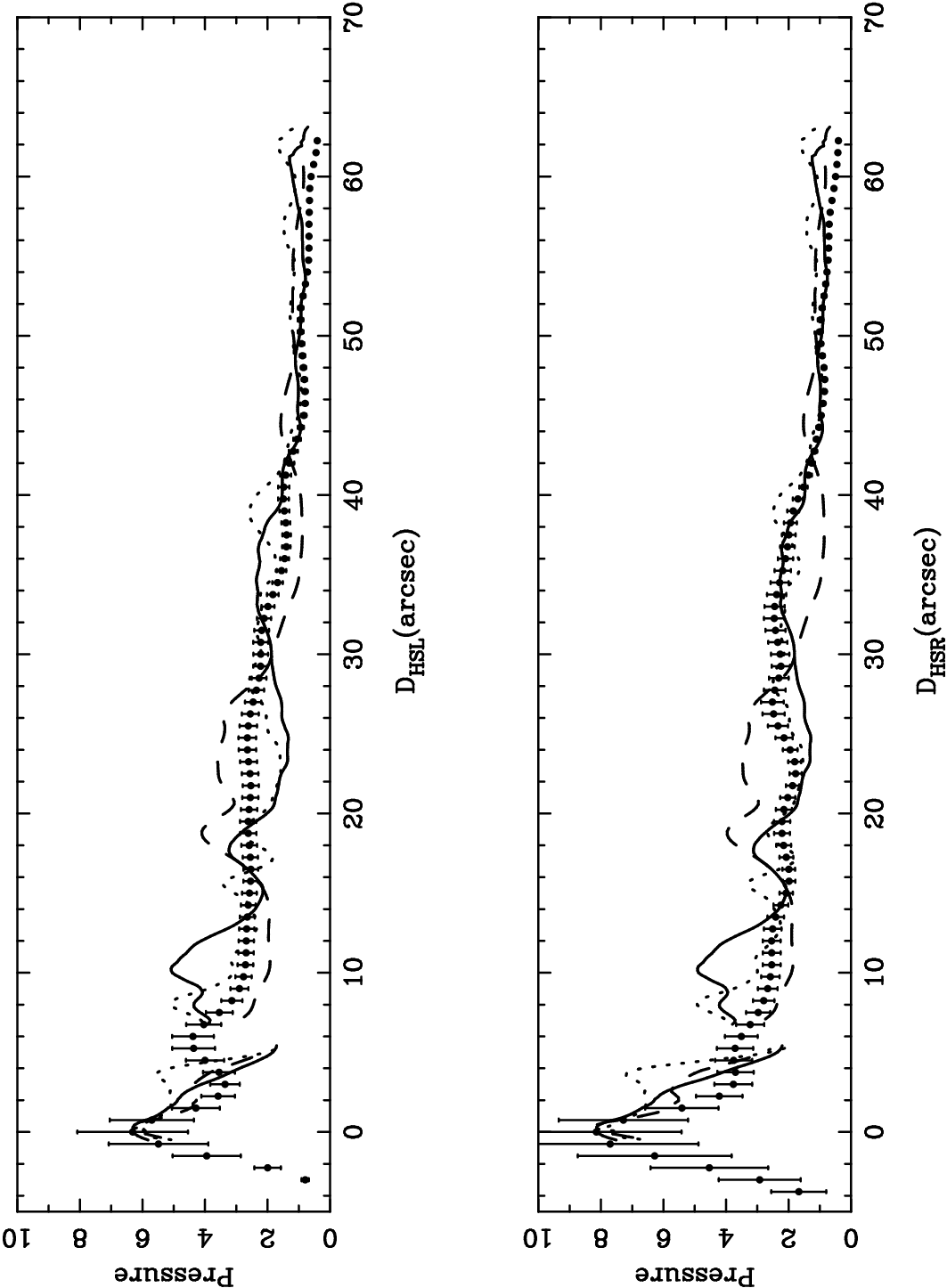}
\figcaption{ Comparison between observed and calculated average thermal
pressure as a function of distance from the hot spot, with the
numerical result scaled to match the empirical result at the 
hot spot peak. The Left
side of the source is shown in the upper panel, and the Right side
is shown in the lower panel. The dashed line, dotted line and full
line correspond to panels (c), (d) and (e) of Fig.
\ref{AradioD01te}, respectively.
 \label{ACyg_Pressure_var} }
\end{figure}
\clearpage

\begin{figure}[ht]
\epsscale{0.8} 
\plotone{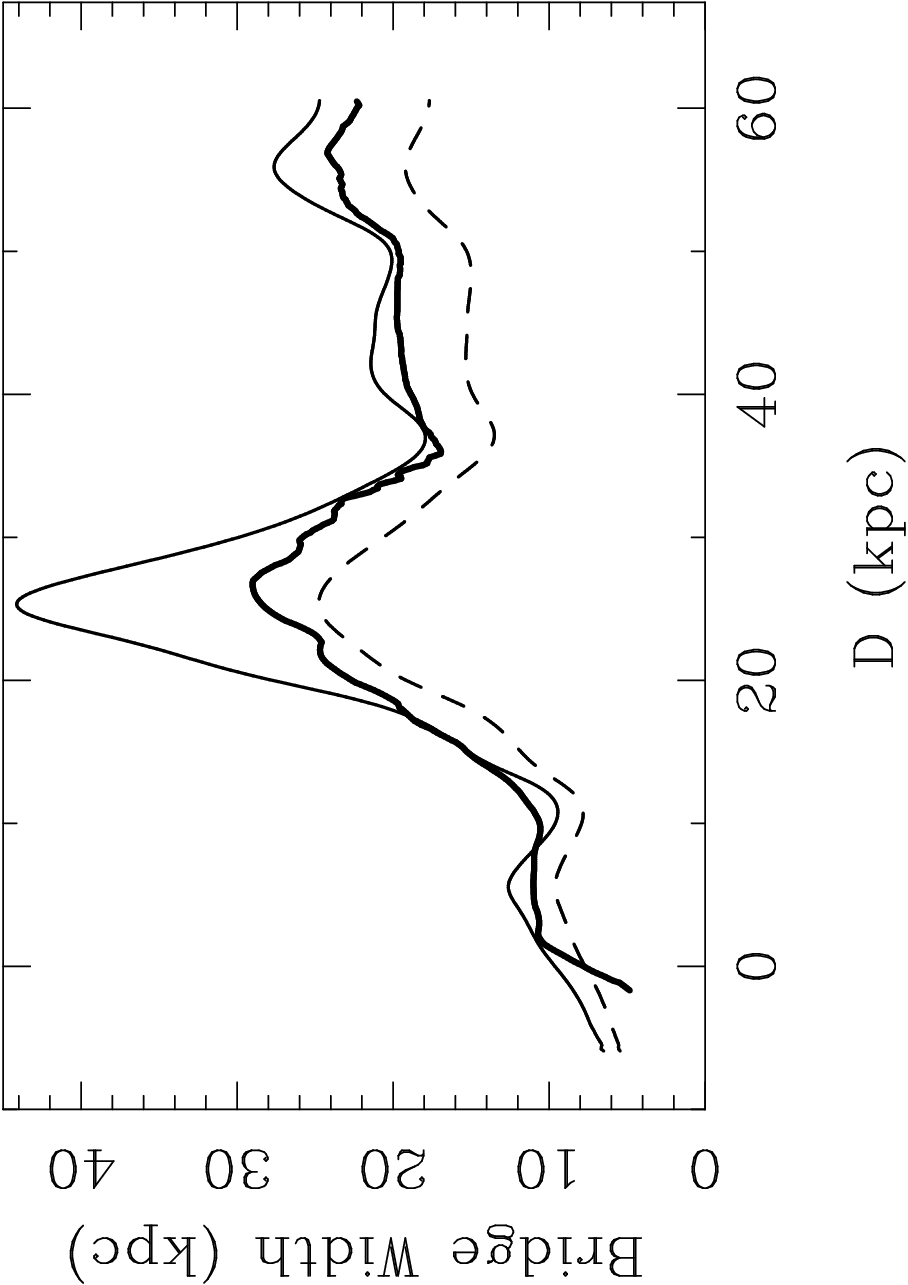}
\figcaption{Bridge width calculated from the Gaussian fit (FWHM)
and the second moment compared with the full width of the cocoon
measured from the simulations as a function of distance from the
hotspot. Thick line is cocoon ($W$), thin line is the Gaussian fit
($W_G$) and dashed line is the second moment ($W_2$).
\label{ACygA-BridgeWD01} }
\end{figure}
\clearpage

\begin{figure}[ht]
\epsscale{0.8} \plotone{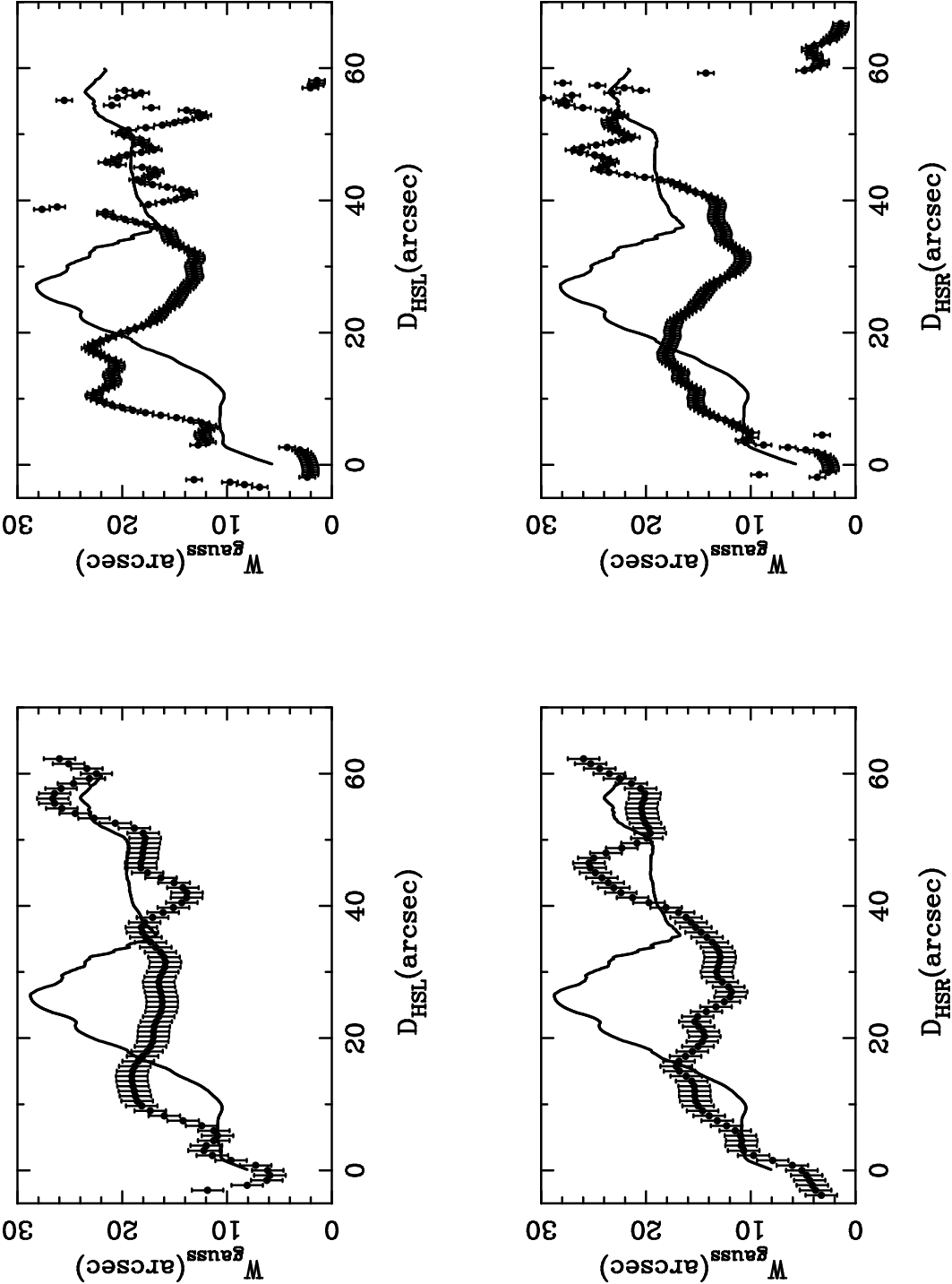}
\figcaption{Comparison between  the FWHM fit to the radio data
with the ``true" width of the simulated source as a function of
distance from the hot spot for 151 MHz (left panels) and 1345 MHz
(right panels). The Left side of the source is shown in the upper
panels, and the Right side is shown in the lower panels.}
\label{ACyg_Width} 
\end{figure}
\clearpage

\begin{figure}[ht]
\epsscale{0.8} 
\plotone{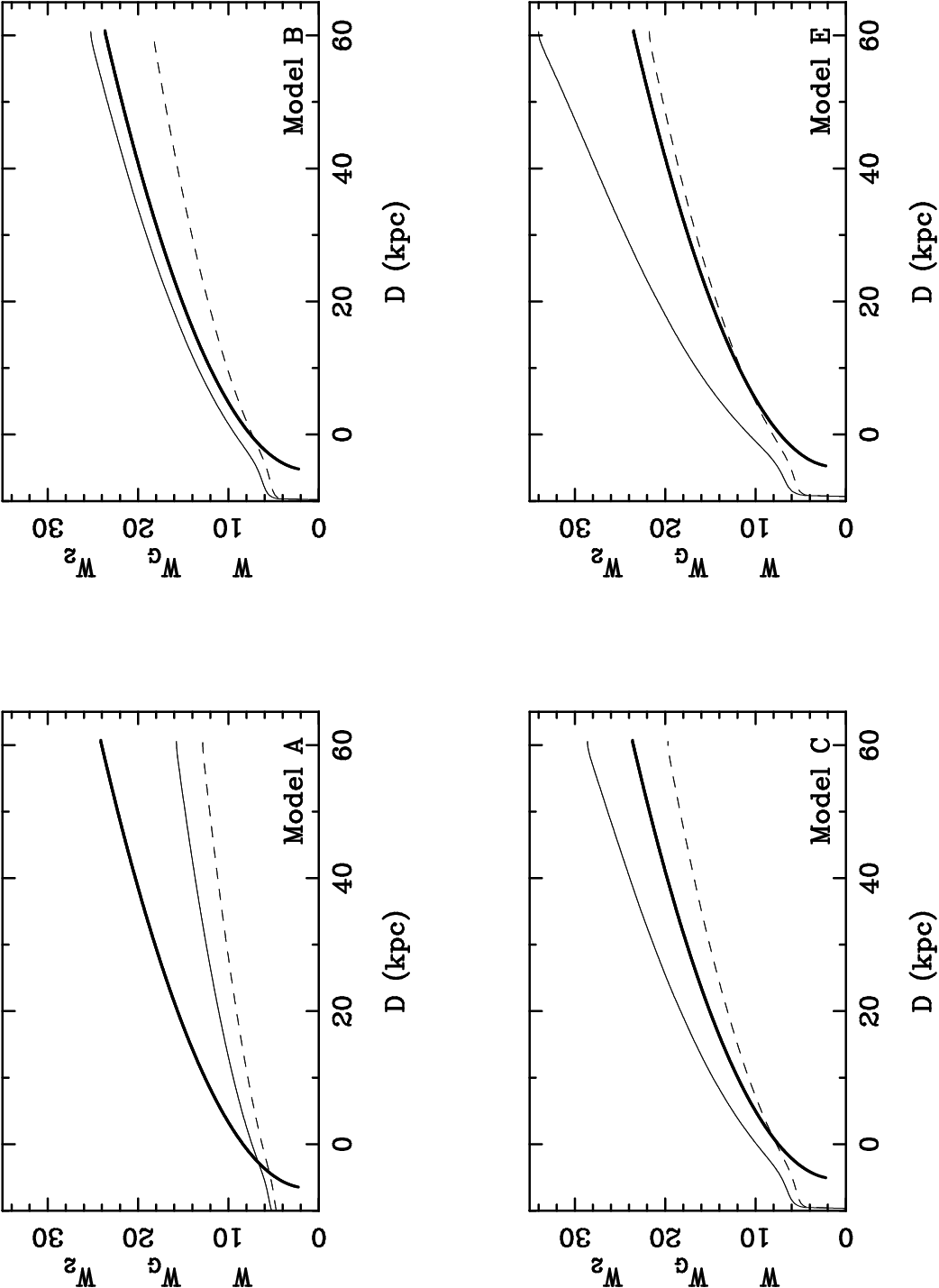}
\figcaption{Bridge width calculated from the Gaussian fit (FWHM)
and the second moment compared with the full width of the model
cocoon as a function of distance from the hotspot. Thick line is
cocoon ($W$), thin line is the Gaussian fit ($W_G$) and dashed
line is the second moment ($W_2$). The four panels represents the
models in Table \ref{zeta}.
\label{ACyg_Width4} }
\end{figure}
\clearpage


\begin{thebibliography}{}


\bibitem[]{} Alexander, P., \&  Leahy, P., 1987, MNRAS, 225, 1


\bibitem[]{} Alexander, P., \& Pooley, G. G. 1996, in Cygnus A - Study of a
        Radio Galaxy, eds. C. L. Carilli \& D. E. Harris, (Cambridge University
        Press, Cambridge), p. 149

\bibitem[]{} Begelman, M. C., \& Cioffi, D. F. 1989, ApJ, 345, L21


\bibitem[]{} Carilli, C. L., Perley, R. A., Dreher, J. W., \& Leahy, J. P.,
1991, ApJ, 383, 554


\bibitem[]{} Carvalho, J. C., \& O'Dea, C. P., 2002a, ApJS, 141, 337


\bibitem[]{} Carvalho, J. C., \& O'Dea, C. P., 2002b, ApJS, 141, 371


\bibitem[]{} Cioffi, D. F., \& Blondin, J. M. 1992, ApJ, 392, 458


\bibitem[]{} Clarke, D. A., Norman, M. L. \& Burns, J. O. 1989, ApJ, 342, 700

\bibitem[]{} Cox, C. I., Gull, S. F., \& Scheuer, P. A. G. 1991, MNRAS,
252, 558

\bibitem[]{} Daly, R. A. 1990, ApJ, 355, 416

\bibitem[]{} Daly, R. A. 1994, ApJ, 426, 38

\bibitem[]{} De Young, D. S. 1980, ApJ, 241, 81

\bibitem[]{} Donahue, M., Daly, R. A., \& Horner, D. J. 2003, ApJ, 584, 643

\bibitem[]{} Eilek, J. A, \& Henriksen, R. N. 1984, ApJ, 277, 820

\bibitem[]{} Eilek, J. A., \& Shore, S. N. 1989, ApJ, 342, 187

\bibitem[]{} Falle, S. A. E. G. 1991, MNRAS, 250, 581

\bibitem[]{} Hardcastle, M. J., Birkinshaw, M, Cameron, R. A.,
Harris, D. E., Looney, L. W., \& Worrall, D. M. 2002, ApJ, 581, 948 

\bibitem[]{} Hardcastle, M. J., Harris, D. E., Worrall, D. M.,
\& Birkinshaw, M. 2004, ApJ, in press (astro-ph/0405516)

\bibitem[]{} Hardee, P. E., \& Norman, M. L. 1990, ApJ, 365, 134

\bibitem[]{} Kardashev, N. S., 1962, Soviet-AJ, 6, 3171

\bibitem[]{} Komissarov, S. S., \& Falle, S. A. E. G. 1997, MNRAS, 288, 833


\bibitem[]{} Komissarov, S. S., \& Falle, S. A. E. G. 1998, MNRAS, 297, 1087


\bibitem[]{} Krause, M. 2003, A\&A, 398, 113


\bibitem[]{} Leahy, J. P. 1991, In Beams and Jets in Astrophysics, ed. P. A.
        Hughes, (Cambridge University Press), 100

\bibitem[]{} Leahy, J. P., Muxlow, T. W. B., \& Stephens, P. W. 1989, MNRAS,
239, 401


\bibitem[]{} Lind K.R., Payne D.G., Meier D.L., \& Blandford R.D. 1989, ApJ,
344, 89


\bibitem[]{} Loken, C., Burns, J. O., Clarke, D. A., \& Norman, M. L. 1992, ApJ,
392, 54


\bibitem[]{} Marti, J. M$^a$., M\"uller, E., Font, J. A., Ibanez, J. M$^a$., \&
Marquina, A. 1997, ApJ, 479, 151


\bibitem[]{} Mioduszewski, A. J., Hughes, P. A., \& Duncan, G. C. 1997, ApJ,
476, 649


\bibitem[]{} Norman, M. L., Smarr, L., Winkler, K.-H., A., \& Smith, M. D.,
1982, A\&A, 113, 285

\bibitem[]{} Pacholczyk, A. G., 1970, Radio Astrophysics (San Francisco: Freeman)


\bibitem[]{} Perley, R. A, \& Taylor, G. B.,  1991, AJ, 101, 1623

\bibitem[]{} Reynolds, C. S., Heinz, S., \& Begelman, M. C., 2002, MNRAS, 332, 271

\bibitem[]{} Rosen, A., Hughes, P. A., Duncan, G. C., \& Hardee, P. E. 1999,
ApJ, 516, 729

\bibitem[]{} Scheuer, P. A. G. 1982, in: Extragalactic Radio Sources, IAU
Symposium No. 97, eds. D. S. Heeschen \& C. M. Wade (D. Reidel), p. 163

\bibitem[]{} Smith, D. A., Wilson, A. S., Arnaud, K. A., Terashima, Y, 
\& Young, A. J. 2002, ApJ, 565, 195

\bibitem[]{} Wan, L., Daly, R.A., \& Guerra, E. J., 2000, ApJ, 544, 671 

\bibitem[]{} Wellman, G.F., Daly, R.A., \& Wan, L. 1997a, ApJ, 480, 79.

\bibitem[]{} Wellman, G.F., Daly, R.A., \& Wan, L. 1997b, ApJ, 480, 96.

\bibitem[]{} Wilson, A. S., Young, A. J., \& Shopbell, P. L. 2000, ApJ, 544, 27

\end{thebibliography}
\end{document}